\documentclass[12pt]{article}

\pdfoutput=1

\usepackage{graphics}
\usepackage{epsfig}
\usepackage{amsfonts}
\usepackage{amssymb}
\usepackage{amsmath}
\usepackage{dsfont}
\usepackage{multirow}
\usepackage{verbatim}
\usepackage{cite}
\usepackage{cancel}
\usepackage{slashed}

\newcommand{\beq}{\begin{equation}}
\newcommand{\eeq}{\end{equation}}
\newcommand{\be}{\begin{equation}}
\newcommand{\ee}{\end{equation}}
\newcommand{\ben}{\begin{displaymath}}
\newcommand{\een}{\end{displaymath}}
\newcommand{\bea}{\begin{eqnarray}}
\newcommand{\eea}{\end{eqnarray}}

\newcommand{\bean}{\begin{eqnarray*}}
\newcommand{\eean}{\end{eqnarray*}}

\DeclareMathAlphabet{\mathpzc}{OT1}{pzc}{m}{it}

\begin{document}
\pagestyle{plain}


\makeatletter \@addtoreset{equation}{section} \makeatother
\renewcommand{\thesection}{\arabic{section}}
\renewcommand{\theequation}{\thesection.\arabic{equation}}
\renewcommand{\thefootnote}{\arabic{footnote}}


\setcounter{page}{1} \setcounter{footnote}{0}


\begin{titlepage}

\begin{flushright}
UUITP-15/14\\
NIKHEF 2014-047\\
\end{flushright}

\bigskip

\begin{center}

\vskip -10mm

{\LARGE \bf KK-monopoles and G-structures \\[4mm] in M-theory/type IIA reductions} \\[6mm]

\vskip 0.5cm

{\bf Ulf Danielsson$^1$, Giuseppe Dibitetto$^1$ \,and\,  Adolfo Guarino$^{2,3}$}\let\thefootnote\relax\footnote{{\tt ulf.danielsson@physics.uu.se,  giuseppe.dibitetto@physics.uu.se, aguarino@nikhef.nl}}\\

\vskip 25pt

{\em $^1$Institutionen f\"or fysik och astronomi, University of Uppsala, \\ Box 803, SE-751 08 Uppsala, Sweden \\[2mm]
$^2$Nikhef, Science Park 105, 1098 XG Amsterdam, The Netherlands \\[2mm]
$^3$Albert Einstein Center for Fundamental Physics, Institute for Theoretical Physics, \\ Bern University, Sidlerstrasse 5, CH–3012 Bern, Switzerland \\}

\vskip 0.8cm

\end{center}

\vskip 1cm

\begin{center}

{\bf ABSTRACT}\\[3ex]

\begin{minipage}{13cm}
\small

We argue that M-theory/massive IIA backgrounds including KK-monopoles are suitably described in the language of G-structures and their intrinsic torsion. To this end, we study classes of minimal supergravity models that admit an interpretation as twisted reductions in which the twist parameters are not restricted to satisfy the Jacobi constraints $\omega\, \omega=0$ required by an ordinary Scherk-Schwarz reduction. We first derive the correspondence between four-dimensional data and torsion classes of the internal space and, then, check the one-to-one correspondence between higher-dimensional and four-dimensional equations of motion. Remarkably, the whole construction holds regardless of the Jacobi constraints, thus shedding light upon the string/M-theory interpretation of (smeared) KK-monopoles.

\end{minipage}

\end{center}

\vfill

\end{titlepage}


\tableofcontents

\newpage

\section{Introduction}
\label{sec:introduction}

Since its very birth, the mechanism of flux compactification has been studied in string and M-theory mainly with the aim of producing
realistic four-dimensional vacuum solutions. Two complementary approaches have been designed to this purpose. The first one, which is often referred to as the \emph{top-down} approach, consists in explicitly performing a dimensional reduction down to lower dimensions on particular backgrounds solving the ten- or eleven-dimensional equations of motion. Such analyses turn out to be quite complicated in general in that they rely on the consistency of the corresponding truncation to the lower-dimensional fields, and need particularly well-suited flux \emph{Ans\"atze} to produce consistently solvable higher-dimensional equations of motion. The second approach is usually called the \emph{bottom-up} approach and focuses on classes of effective (supersymmetric) field theory descriptions in four dimensions, which are generically obtained by certain compactifications of string or M-theory. Within these supergravities, one has the advantage of treating the dynamics of the moduli by simply studying critical points of an effective potential induced by the presence of fluxes. In this way, it becomes exceptionally straightforward to analyse the full set of moduli and not only those ones that can be understood as perturbations of the internal metric.   

Going back to the \emph{top-down} approach, classes of compactifications which have been extensively studied in the literature are those on internal manifolds with G-structure. See \emph{e.g.} refs~\cite{Gauntlett:2003cy,Grimm:2004ua,Grimm:2005fa,House:2005yc,Aldazabal:2007sn,Caviezel:2008ik,Caviezel:2008tf,Danielsson:2009ff,Danielsson:2010bc,Koerber:2010rn,Danielsson:2011au,Danielsson:2012et} for the case of type IIA reductions on $\textrm{SU}(3)$-structures, and refs~\cite{Bilal:2001an,Dall'Agata:2003ir,Martelli:2003ki,Behrndt:2004bh,House:2004pm,Lambert:2005sh} for the case of M-theory on $\textrm{G}_2$-structures. In particular, in ref.~\cite{Danielsson:2010bc} the $\textrm{SU}(3)$-structure defining the internal geometry was also used to construct the so-called \emph{universal Ansatz} for the gauge fluxes. By such a procedure, one is guaranteed to obtain a stress-energy tensor consistent with the form of the Einstein tensor thus making it possible to solve the ten-dimensional equations of motion without posing any extra constraints.

On the other hand, following the \emph{bottom-up} approach, lots of progress has been made in the literature in the last decade especially in the context of certain minimal $\mathcal{N}=1$ supergravities called STU-models. They can be generically obtained by performing a \textit{twisted} orbifold compactification of type II string theories or M-theory. See \emph{e.g.} refs~\cite{Derendinger:2004jn,Dall'Agata:2009gv} for the case of type IIA reductions on $T^{6}/\mathbb{Z}_{2}^{2}$ and refs~\cite{Dall'Agata:2005fm,Derendinger:2014wwa} for the case of M-theory on $T^{7}/\mathbb{Z}_{2}^{3}$. These twisted reductions are proven to be consistent provided that certain \emph{Jacobi constraints} of the form \cite{Scherk:1979zr}
\beq
\label{ww=0_intro}
\omega \, \omega \,\,=\,\, 0 \ ,
\eeq
are satisfied by the twist parateters $\,\omega\,$. However, from the perspective of $\,\mathcal{N}=1\,$ supergravity, these constraints may seem a bit artificial in the sense that minimal supersymmetry would easily allow for their relaxation. Such a possibility has been considered in ref.~\cite{Villadoro:2007yq}, where it was argued that (smeared) KK-monopoles can be viewed as sources to the r.h.s. of the Jacobi constraints
\beq
\label{wwneq0_intro}
\omega \, \omega \,\,\neq \,\, 0 \,\, \Rightarrow \,\, \textrm{KK-monopoles} \ .
\eeq
Some work in this direction has been done in refs~\cite{Gaillard:2009kz,McOrist:2012yc,Andriot:2014uda}, where M-theory backgrounds with KK-monopoles and their relation to type IIA are analysed with several complementary motivations.

The aim of this work is to interpret twisted $\,T^{6}/\mathbb{Z}_{2}^{2}\,$ and $\,T^{7}/\mathbb{Z}_{2}^{3}\,$ reductions as reductions on SU(3)- and G$_2$-structure manifolds, respectively, and establish a solid correspondence between twist parameters (a.k.a. metric fluxes) and torsion classes of the corresponding G-structure. We want to make use of the above correspondence in order to construct an explicit uplift of the STU-models to M-theory or type IIA and provide an interpretation in terms of G-structure compactifications, \textit{regardless} of whether or not the Jacobi constraints are satisfied. This will justify and corroborate the validity of the \emph{bottom-up} approach, at least as far as STU-models are concerned. In addition, it will also shed new light on the nature of (smeared) KK-monopoles by giving them a natural and geometric interpretation: unlike for M-branes or D-branes, the KK-monopoles are secretly built-in within the bulk action of eleven-dimensional and massive type IIA supergravity. As a result, they can be nicely described using the framework of G-structures and their intrinsic torsion. 

The paper is organised as follows. In section~\ref{sec:G-structures}, we review relevant facts regarding M-theory reductions on manifolds with $\textrm{G}_2$-structure as well as massive type IIA reductions on manifolds with $\textrm{SU}(3)$-structure. In section~\ref{sec:Twisted orbifolds} we revisit the STU-models obtained as particular twisted orbifold reductions of either M-theory or massive type IIA string theory and establish the connection to G$_{2}$- and SU(3)-structures and their intrinsic torsion. We compare the two types of reductions in the context of $\textrm{SU}(3)$-structures of $7d$ \textit{vs} $6d$ manifolds, comment on the non-geometric type IIA interpretation of some M-theory reductions and discuss the relaxation of the Jacobi constraints and how this fact is interpreted as adding KK-monopoles as sources. In section~\ref{sec:upliftings} we construct the massive type IIA/M-theory uplift of the corresponding STU-models by studying the ten-/eleven-dimensional equations of motion and showing their one-to-one correspondence with the four-dimensional supergravity equations of motion coming from varying the effective scalar potential of the STU-model. Interestingly, such a correspondence works regardless of the Jacobi constraints, namely, whether or not smeared KK-monopoles are included in the construction. We conclude with a discussion of the results and pose some remaining issues which might open up new possible developments. A summary of conventions concerning the geometry and topology of the twisted $\,T^{6}/\mathbb{Z}_{2}^{2}\,$ and $\,T^{7}/\mathbb{Z}_{2}^{3}\,$ orbifolds is presented in the appendix~\ref{sec:app1}.

\section{M-theory/Type IIA on $\textrm{G}$-structure manifolds}
\label{sec:G-structures}

In this section we review a class of orbifold reductions of M-theory and \emph{massive} type IIA strings on twisted tori with gauge fluxes and their corresponding four dimensional ($4d$) supergravity effective descriptions as STU-models. Furthermore, we will respectively connect them to reductions on seven dimensional ($7d$) $\textrm{G}_{2}$-structure and six dimensional ($6d$) $\textrm{SU}(3)$-structure manifolds.

\subsection{M-theory on a $\textrm{G}_{2}$-manifolds $X_{7}$ with fluxes}
\label{sec:M-theory}

We start with a discussion of the $4d$ effective supergravities coming from reductions of $11d$ supergravity on G$_{2}$-structure manifolds with fluxes.

\subsubsection{G$_2$-structure manifolds}

A seven-dimensional manifold $X_{7}$ with a G$_{2}$-structure \cite{Friedrich:2002,Bryant:2003} is specified in terms of a G$_{2}$ invariant three-form $\Phi_{(3)}$ or, equivalently, in terms of a covariantly constant spinor $\eta$ such that $\,\Phi_{ABC}\propto \eta^{\dagger} \gamma_{ABC} \,\eta\,$ with $\,A=,1,...,7$. The presence of such G$_{2}$ invariant objects can be inferred from the decomposition of the corresponding SO(7) representations under the maximal $\textrm{G}_{2} \subset \textrm{SO}(7)$ subgroup
\be
\label{branching_SO(7)}
\begin{array}{cclc}
\textrm{SO}(7) & \supset & \textrm{G}_{2} & \\[2mm]
\textbf{7} & \rightarrow & \textbf{7} & , \\[2mm]
\textbf{8} & \rightarrow & \textbf{1} \oplus \textbf{7}& , \\[2mm]
\textbf{21} & \rightarrow & \textbf{7} \oplus \textbf{14}& , \\[2mm]
\textbf{35} & \rightarrow & \textbf{1} \oplus \textbf{7} \oplus \textbf{27} & .
\end{array}
\ee
The invariant three-form $\Phi_{(3)}$ thus corresponds to the singlet appearing in the decomposition of the $\textbf{35}$. The failure in the closure of $\Phi_{(3)}$ is understood as the presence of a non-vanishing torsion
\be
\begin{array}{lclclc}
{T_{AB}}^{C} & \in & \Lambda^{2}(X_{7}) \ \otimes \ \Lambda^{1}(X_{7})\ & = & ( \textbf{7} \ \oplus \ 
\underbrace{\xcancel{\, \textbf{14} \,}}_{\textrm{adj}(\textrm{G}_{2})}) \ \otimes \ \textbf{7} & ,
\end{array}
\ee
which  splits into a set of torsion classes, namely different G$_{2}$ representations, given by
\be
\begin{array}{lclclc}
{T_{AB}}^{C} & \rightarrow & \underbrace{\textbf{1}}_{\tau_{0}} \ \oplus \ \underbrace{\textbf{7}}_{\tau_{1}} \ \oplus \ 
\underbrace{\textbf{14}}_{\tau_{2}} \ \oplus \ 
\underbrace{\textbf{27}}_{\tau_{3}} & ,
\end{array}
\ee
satisfying the linear relations
\be
\Phi_{(3)} \wedge \star_{7d} \tau_{3} = 0
\hspace{10mm} \textrm{ and } \hspace{10mm}
\Phi_{(3)} \lrcorner \star_{7d} \tau_{3} = 0 \ .
\ee
As anticipated, the above torsion classes act as sources in the r.h.s of the closure relations for the invariant three-form $\Phi_{(3)}$ and its $7d$ dual four-form $\star_{7d} \Phi_{(3)}$. These are given by
\be
\label{dPhi}
\begin{array}{lclc}
d\Phi_{(3)} & = & \tau_{0} \,\, \star_{7d} \Phi_{(3)} \,\,+ \,\, 3 \, \tau_{1} \, \wedge \Phi_{(3)} \,\,+ \,\,\star_{7d} \tau_{3}  & , \\[2mm]
d\star_{7d} \Phi_{(3)} & = & 4 \, \tau_{1}  \wedge \star_{7d} \Phi_{(3)} \,\, + \,\, \tau_{2} \wedge \Phi_{(3)}   & .
\end{array}
\ee
Later on we will restrict to the case of vanishing $\tau_{1}\,=\,\tau_{2}\,=\,0$ which corresponds to co-calibrated G$_{2}$ structures. These include the case of the $X_{7}=T^{7}/\mathbb{Z}_{2}^{3}$ orbifold we will investigate in this work.

\subsubsection{Ricci scalar and scalar potential}

The $7d$ metric $g_{AB}$ of the G$_{2}$-manifold can be constructed from the invariant form $\Phi_{(3)}$ using the standard formula
\be
\label{g7_metric}
g^{(7)}_{AB} = \det(h_{AB})^{-1/9} \, h_{AB}
\hspace{10mm} \textrm{with} \hspace{10mm}
h_{AB} = \frac{1}{144} \, \epsilon^{C_{1}...C_{7}} \, \Phi_{A C_{1} C_{2}} \, \Phi_{B C_{3} C_{4}} \, \Phi_{C_{5} C_{6} C_{7}} \ .
\ee
The associated Ricci scalar can be expressed in terms of the torsion classes entering (\ref{dPhi}). The result is given by
\be
\label{R7}
\mathcal{R}^{(7)}  = -12 \, \star_{7d} d \star_{7d} \tau_{1}  \,+\,  \frac{21}{8} \, \tau_{0}^{2} \, + 30 \, |\tau_{1}|^{2} \, - \, \frac{1}{2} \, |\tau_{2}|^{2}  \, - \, \frac{1}{2} \, |\tau_{3}|^{2} \ ,
\ee
where $\,|\tau_{p}^{2}| \equiv \frac{1}{p!} \, \tau_{A_{1}...A_{p}} \tau^{A_{1}...A_{p}}\,$ and where seven-dimensional indices are raised using the inverse of the metric $g^{(7)}_{AB}$ introduced in (\ref{g7_metric}).

The $7d$ Ricci scalar (\ref{R7}) becomes (part of) the scalar potential upon reduction of the $11d$ Ricci scalar 
\be
S_{11d} \supset \int d^{11}x \, \sqrt{g^{(11)}}\, \mathcal{R}^{(11)} \ .
\ee
Taking the $11d$ metric to be of the form
\be
ds_{(11)}^{2} = \tau^{-2} \, ds_{(4)}^{2} + ds_{(7)}^{2} \ ,
\ee
requires the four-dimensional dilaton to be identified as $\tau^{2}=\sqrt{g^{(7)}}$ in order to recover the Einstein frame in four dimensions. This is then compatible with a four-dimensional action of the form
\be
S_{4d} \supset \int d^{4}x \, \sqrt{g^{(4)}}\, \Big( \mathcal{R}^{(4)} + \frac{1}{\sqrt{g^{(7)}}} \, \mathcal{R}^{(7)}\Big) \ ,
\ee
which results in the appearance of a scalar potential due to the internal geometry of the form
\be
\label{VRicci_M-theory}
V_{\textrm{G$_{2}$}} = -\frac{1}{\sqrt{g^{(7)}}} \, \mathcal{R}^{(7)} \ .
\ee
We will verify the above relation latter on for the case of the the $X_{7}=T^{7}/\mathbb{Z}_{2}^3$ orbifold for which $\tau_{1} = \tau_{2}=0$.

\subsubsection{$\mathcal{N}=1$ effective action and flux-induced superpotential }

Because of the singlet in the decomposition of the $\textbf{8}$ in (\ref{branching_SO(7)}), reductions of M-theory on G$_{2}$-manifolds with fluxes produce $\mathcal{N}=1$ effective supergravities in $4d$. The M-theory flux-induced superpotential is given by \cite{House:2004pm,Dall'Agata:2005fm}
\be
\label{Potential_M-theory}
\mathcal{W}_{\textrm{M-theory}} = \frac{1}{4} \int_{X_{7}} G_{(7)} + \frac{1}{4} \int_{X_{7}} (C_{(3)} + i \Phi_{(3)}) \wedge \left[ G_{(4)}+\frac{1}{2} d (C_{(3)} + i \Phi_{(3)})\right] \ ,
\ee
where, for the twisted orbifold reductions we will consider in this work, $d$ is the $7d$ \textit{twisted} derivative operator $d=\partial + \omega$ acting on a generic $p$-form $T_{(p)}$ as 
\beq
(dT)_{A_{1}...A_{p+1}}=\partial_{[A_{1}} \, T_{A_{2}...A_{p+1}]} \, - \,  {\omega_{[A_{1}A_{2}}}^{B} \, T_{|B|A_{3}...A_{p+1}]} \ ,
\eeq
with $\,{\omega_{AB}}^{C}\,$ being the $7d$ twist parameters (metric fluxes). $C_{(3)}$ is the three-form gauge potential of $11d$ supergravity and $G_{(4)}$ its associated background flux along the internal directions. In addition, $G_{(7)}$ corresponds to the dual of a background flux entirely along the external directions, \textit{i.e.} a Freund-Rubin parameter \cite{Freund:1980xh}. Having non-vanishing $G_{(7)}\neq0$ proved to be a necessary ingredient to fully stabilise moduli in the $X_{7}=T^{7}/\mathbb{Z}_{2}^3$ reductions of refs~\cite{Dall'Agata:2009gv,Derendinger:2014wwa}.

\subsection{Massive IIA on an $\textrm{SU}(3)$-manifold $X_{6}$ with fluxes}
\label{sec:IIA-theory}

Let us now discuss the $4d$ effective supergravities arising upon reduction of \textit{massive} type IIA supergravity on SU(3)-structure manifolds with fluxes.

\subsubsection{SU(3)-structure manifolds}

A six-dimensional manifold $X_{6}$ with $\textrm{SU}(3)$-structure is characterised by the presence of two globally defined and $\textrm{SU}(3)$-invariant fundamental forms -- a real 2-form $J$ and a holomorphic
3-form $\Omega$ -- defining an interpolation between a complex and a symplectic structure. 
By decomposing the 2- and (anti-)self-dual 3-form representations of $\textrm{SO}(6)$ w.r.t. its $\textrm{SU}(3)$ maximal subgroup, one indeed finds 
\be
\label{branching_SO(6)}
\begin{array}{cclc}
\textrm{SO}(6) & \supset & \textrm{SU}(3) & \\[2mm]
\textbf{8} & \rightarrow & \textbf{1} \ \oplus \ \textbf{1} \ \oplus \ \textbf{3} \ \oplus \ \bar{\textbf{3}}  & , \\[2mm]
\textbf{15} & \rightarrow & \textbf{1} \ \oplus \ \textbf{3} \ \oplus \ \bar{\textbf{3}} \ \oplus \ \textbf{8} & , \\[2mm]
\textbf{10} & \rightarrow & \textbf{1}\ \oplus \ \textbf{3} \ \oplus \ \textbf{6} & , \\[2mm]
\textbf{10}^{\prime} & \rightarrow & \textbf{1} \ \oplus \ \bar{\textbf{3}} \ \oplus \ \bar{\textbf{6}} & , 
\end{array}
\ee
featuring the three singlets corresponding to $J$, $\Omega$ and $\bar{\Omega}$ respectively. Besides the above topological constraint, the $\textrm{SU}(3)$-structure (together with supersymmetry) requires a set of special differential conditions which select only some allowed $\textrm{SU}(3)$ irrep's
inside the expression of the exterior derivatives of the above fundamental forms. Such irrep's are usually referred to as torsion classes and can be obtained by decomposing the most general metric connection
${T_{mn}}^{p}$ into $\textrm{SU}(3)$ pieces
\be
\begin{array}{lclclc}
{T_{mn}}^{p} & \in & \Lambda^{2}(X_{6}) \ \otimes \ \Lambda^{1}(X_{6})\ & = & (\textbf{1} \ \oplus \ \textbf{3} \ \oplus \ \bar{\textbf{3}} \ \oplus \ 
\underbrace{\xcancel{\, \textbf{8} \,}}_{\textrm{adj}(\textrm{SU}(3))}) \ \otimes \ (\textbf{3} \ \oplus \ \bar{\textbf{3}}) & ,
\end{array}
\ee
where the contribution coming from the adjoint representation of $\textrm{SU}(3)$ has been crossed out since it drops out whenever acting on invaraint forms like $J$ and $\Omega$ \cite{Grana:2005jc}. This procedure yields
\be
\begin{array}{lclclc}
{T_{mn}}^{p} & \rightarrow & \underbrace{(\textbf{1} \ \oplus \ \textbf{1})}_{W_{1}} \ \oplus \ \underbrace{(\textbf{8} \ \oplus \ \textbf{8})}_{W_{2}} \ \oplus \ 
\underbrace{(\textbf{6} \ \oplus \ \bar{\textbf{6}})}_{W_{3}} \ \oplus \ 
\underbrace{2 \, \times \, (\textbf{3} \ \oplus \ \bar{\textbf{3}})}_{\left(W_{4}, \ W_{5}\right)} & ,
\end{array}
\ee
where $W_{1}$ is a complex 0-form, $W_{2}$ is a complex primitive 2-form, \emph{i.e.} such that
\be\label{primitivity1}
W_{2} \, \wedge \, J \, \wedge \, J \, = \, 0 \ ,
\ee
$W_{3}$ is a real primitive 3-form, \emph{i.e.} such that
\be\label{primitivity2}
W_{3} \, \wedge \, \Omega \, = \, 0 \ ,
\ee
and, finally, $W_{4}$ and $W_{5}$ are real 1-forms. The full expression of the exterior derivatives of the fundamental forms in terms of the torsion classes reads
\be
\label{dJdOmega}
\begin{array}{lclc}
dJ & = & \dfrac{3}{2} \, \textrm{Im}(\bar{W}_{1}\,\Omega) \ + \ W_{4} \, \wedge \, J \ + \ W_{3} & , \\[2mm]
d\Omega & = & W_{1} \, J \, \wedge \, J \ + \ W_{2} \, \wedge \, J \ + \ \bar{W}_{5} \, \wedge \, \Omega & .
\end{array}
\ee
We shall in the following restrict ourselves to the case $W_{4}\,=\,W_{5}\,=\,0$, which certainly includes the example of $X_{6}=T^{6}/\mathbb{Z}_{2}^2$ that we want to make contact with, as well as any other manifold $X_{6}$ without 1- and 5-cycles.

\subsubsection{Ricci scalar and scalar potential}

In terms of the fundamental forms, one can subsequently introduce a metric on $X_{6}$. The intermediate step is defining the quantity \cite{Danielsson:2011au}
\be
\label{I-involution}
{I_{m}}^{n} \ \equiv \ \lambda \, \epsilon^{m_{1}m_{2}m_{3}m_{4}m_{5}n} \, \left(\Omega_{R}\right)_{mm_{1}m_{2}} \, \left(\Omega_{R}\right)_{m_{3}m_{4}m_{5}} \ ,
\ee
where $\,\Omega_{R} \equiv \textrm{Re}(\Omega)\,$ , $\,\Omega_{I} \equiv \textrm{Im}(\Omega)\,$ and $\lambda$ is a moduli-dependent quantity fixing the correct normalisation of $I$ to $I^{2} \, \overset{!}{=} \, - \mathds{1}_{6}$. As a consequence, the metric is defined as
\be
\label{metric6(J,I)}
g^{(6)}_{mn} \ \equiv \ J_{mp} \, {I_{n}}^{p} \ .
\ee

The Ricci scalar $\mathcal{R}^{(6)}$ for such six-dimensional $\textrm{SU}(3)$-structure manifolds is then expressed as a function of the torsion classes via \cite{BV:2007}
\be
\label{R6}
\mathcal{R}^{(6)} \ = \ 2 \star_{6d}d\star_{6d} \left(W_{4}+W_{5}\right) \, + \, \frac{15}{2} \, \left|W_{1}\right|^{2} \, - \, \frac{1}{2} \, \left|W_{2}\right|^{2} \, - \, \frac{1}{2} \, \left|W_{3}\right|^{2} \, - \, \left|W_{4}\right|^{2} \, + \, 4 W_{4}\cdot W_{5}\ ,
\ee
where $\left|W_{1}\right|^{2} \ \equiv \ W_{1} \, \bar{W}_{1}$, $\left|W_{2}\right|^{2} \ \equiv \ \frac{1}{2!}\, \left(W_{2}\right)_{mn} \, \left(\bar{W}_{2}\right)^{mn}$, $\left|W_{3}\right|^{2} \ \equiv \ \frac{1}{3!}\, \left(W_{3}\right)_{mnp} \, \left(W_{3}\right)^{mnp}$, $\left|W_{4,5}\right|^{2}\ \equiv \ \left(W_{4,5}\right)_{m} \, \left(W_{4,5}\right)^{m}$, $W_{4}\cdot W_{5} \ \equiv \ \left(W_{4}\right)_{m} \, \left(W_{5}\right)^{m}$ and all six-dimensional indices are raised and lowered with the metric \eqref{metric6(J,I)}.

As happened before, the $6d$ Ricci scalar (\ref{R6}) becomes (part of) the scalar potential upon reduction of the $10d$ Ricci scalar in the string frame
\be
S_{10d} \supset \int d^{10}x \, \sqrt{g^{(10)}}\, e^{-2 \phi} \, \mathcal{R}^{(10)} \ .
\ee
Taking the $10d$ metric to be of the form
\be
ds_{(10)}^{2} = \tau^{-2} \, ds_{(4)}^{2} + ds_{(6)}^{2} \ ,
\ee
requires the four-dimensional dilaton $\phi_{4}$ to be identified as $\tau^{2}=e^{-2 \phi} \,\sqrt{g^{(6)}} \equiv e^{-2 \phi_{4}}$ in order to recover the Einstein frame in four dimensions. This is then compatible with a four-dimensional action of the form
\be
S_{4d} \supset \int d^{4}x \, \sqrt{g^{(4)}}\, \Big( \mathcal{R}^{(4)} + \frac{e^{2 \phi}}{\sqrt{g^{(6)}}} \, \mathcal{R}^{(6)}\Big) \ ,
\ee
which results in the appearance of a scalar potential due to the internal geometry of the form
\be
\label{VRicci_IIA}
V_{\textrm{SU(3)}} = -\frac{e^{2 \phi}}{\sqrt{g^{(6)}}} \, \mathcal{R}^{(6)}\ .
\ee
We will verify the above relation latter on for the case of the the $X_{6}=T^{6}/\mathbb{Z}_{2}^2$ orbifold for which $W_{4}=W_{5}=0$.

\subsubsection{$\mathcal{N}=1$ orientifolds and flux-induced superpotential }

The presence of two singlets in the decomposition of the $\textbf{8}$ in (\ref{branching_SO(6)}) indicates that reductions of type IIA supergravity in SU(3)-manifolds produce $\mathcal{N}=2$ effective supergravities in $4d$. Further applying an orientifold projection, the resulting $\mathcal{N}=1$ supergravity is specified in terms of the flux-induced superpotential \cite{Derendinger:2004jn,Villadoro:2005cu}
\be
\label{Potential_IIA}
\mathcal{W}_{\textrm{IIA}} = \int_{X_{6}} e^{J_{c}} \wedge F + \int_{X_{6}} \Omega_{c} \wedge (H_{(3)} + dJ_{c}) \ .
\ee
In the case of twisted orbifold reductions, the operator $d$ is the $6d$ \textit{twisted} derivative $d=\partial + \omega$ acting on a generic $p$-form $T_{(p)}$ as 
\beq
(dT)_{m_{1}...m_{p+1}}=\partial_{[m_{1}} \, T_{m_{2}...m_{p+1}]} \, - \,  {\omega_{[m_{1}m_{2}}}^{n} \, T_{|n|m_{3}...m_{p+1}]} \ ,
\eeq
with $\,{\omega_{mn}}^{p}\,$ being the $6d$ twist parameters (metric fluxes). $J_{c}\,$ is the complexified K\"aler form $\,J_{c}=B + i \, J\,$ containing the $B$-field and $\,\Omega_{c}\,$ is the complex three-form $\,{\Omega_{c}=C_{(3)} + i \,e^{-\phi} \, \Omega_{R}}\,$ including the R-R potential $C_{(3)}$ and the dilaton field $\phi$ of the $10d$ type IIA supergravity. The above superpotential is induced by the NS-NS background flux $H_{(3)}$ as well as by the R-R background flux $F=\sum_{p} F_{(p)}$, where $F$ denotes the formal sum of $p$-form fluxes $(p=0,2,4,6)$ pairing with the appropriate term in the expansion $e^{J_{c}}=1+J_{c}+...\,$. Similarly to the M-theory case, a background flux $F_{(4)}$ along the external space is traded by a purely internal $F_{(6)}$ flux in (\ref{Potential_IIA}). Finally, the term of the form $\,F_{(0)} \, J_{c} \wedge J_{c} \wedge J_{c}\,$ in (\ref{Potential_IIA}) descends from the Romans mass deformation in the $10d$ type IIA supergravity and turned out to be crucial for moduli stabilisation in the $X_{6}=T^{6}/\mathbb{Z}_{2}^2$ orientifold reductions of refs~\cite{Derendinger:2004jn,Dibitetto:2011gm,Dibitetto:2012ia}.

\section{Twisted orbifolds}
\label{sec:Twisted orbifolds}

The aim of this section is to provide explicit examples of $X_{7}$ and $X_{6}$ manifolds with G$_{2}$- and SU(3)-structure respectively and investigate their connection to flux compactifications of M-theory/Type IIA.

\subsection{STU-models from M-theory/Type IIA}

Twisted orbifolds provide simple examples of manifolds with G$_{2}$- and SU(3)-structure which are easy to handle. In particular we will focus on the $X_{7}=T^{7}/\mathbb{Z}_{2}^3$ orbifold in the case of M-theory reductions and ${X_{6}=T^{6}/\mathbb{Z}_{2}^2}$ for orientifolds of type IIA reductions. Twisting an orbifold amounts to introduce a constant metric $\omega$-flux such that the left-invariant forms $\eta^{A}$ ($\eta^{m}$) globally defined in $X_{7}$ ($X_{6}$) satisfy the Maurer-Cartan equations
\be
\label{Maurer-Cartan}
\begin{array}{llll}
\textrm{Twisted $\,X_{7}$ :} & \hspace{10mm} d \eta^{A} \,+\, \frac{1}{2} \, {\omega_{BC}}^{A} \, \eta^{B} \, \wedge \, \eta^{C} \, = \, 0 & \hspace{8mm} \textrm{with} \hspace{8mm} A=1,...,7 \\[2mm]
\textrm{Twisted $\,X_{6}$ :} & \hspace{10mm} d \eta^{m} \,+\, \frac{1}{2} \, {\omega_{np}}^{m} \, \eta^{n} \, \wedge \, \eta^{p} \, = \, 0 & \hspace{8mm} \textrm{with} \hspace{8mm} m=1,...,6
\end{array}
\ee 
compatible with the orbifold symmetries. In addition to the non-trivial geometry, it is also possible to turn on background fluxes along the internal space for the set of M-Theory/Type IIA gauge potentials in the reduction scheme. This has to be done again respecting the orbifold symmetries. The sets of gauge fluxes consist of
\begin{itemize}
\item[$\circ$] M-theory : $G_{(4)}$ and $G_{(7)}$ background fluxes 
\item[$\circ$] Type IIA : $H_{(3)}$ (NS-NS) and $F_{(p)}$ with $\,p=0,2,4,6\,$ (R-R) background fluxes 
\end{itemize}
and, together, metric and gauge fluxes induce the holomorphic superpotentials in (\ref{Potential_M-theory}) and (\ref{Potential_IIA}). For the twisted orbifolds $X_{7}=T^{7}/\mathbb{Z}_{2}^3$ and $X_{6}=T^{6}/\mathbb{Z}_{2}^2$ we are considering in this work, the reduction gives rise to an $\mathcal{N}=1$ supergravity\footnote{The discrete orbifold action reduces the amount of supersymmetry in the effective action to four supercharges ($\mathcal{N}=1$) in the M-theory case and eight supercharges ($\mathcal{N}=2$) in the case of Type IIA reductions. Modding out the latter by an extra $\mathbb{Z}_{2}$ orientifold action further reduces to four supercharges ($\mathcal{N}=1$).}, more concretely, to a so-called STU-model. The M-theory/Type IIA flux content compatible with the orbifold symmetries is summarised in Table~\ref{Table:Fluxes_M/IIA}.

\begin{table}[t!]
\renewcommand{\arraystretch}{1.35}
\begin{center}
\begin{tabular}{|c|c|c|c|c|}
\hline
STU coupling & M-theory picture & Type IIA picture & Fluxes & dof's \\
\hline
\hline
$U_{I}  \, T_{J}$ ; $S \, U_{I}$& ${\omega_{b c}}^{a}$ , ${\omega_{a j}}^{k}$ ,  ${\omega_{ka}}^{j}$ ; ${\omega_{jk}}^{a}$ & ${\omega_{b c}}^{a}$ , ${\omega_{a j}}^{k}$ , ${\omega_{ka}}^{j}$ ; ${\omega_{jk}}^{a}$ & $\mathcal{C}_{1}^{(IJ)}$ ; $ b_{1}^{\,(I)}$ &  $9+3$ \\
\hline
$U_{J}  \, U_{K}$ & $-{\omega_{ai}}^{7}$ & $ F_{ai}$ & $a_{2}^{\,(I)}$ & $3$  \\
\hline
$S \, T_{I}$ & $-{\omega_{7i}}^{a}$ & non-geometric & $d_{0}^{\,(I)}$ &  $3$ \\
\hline
$T_{J}  \, T_{K}$ & $-{\omega_{a7}}^{i}$ & non-geometric &  $c_{3}'^{\,(I)}$ &  $3$\\
\hline
\hline
$U_{I}$ & $- G_{aibj}$ & $-F_{aibj}$ & $a_{1}^{\,(I)}$ &  $3$ \\
\hline
$S$ & $G_{ijk7}$ & $H_{ijk}$ & $b_{0}$ &  $1$\\
\hline
$T_{I}$ & $G_{ibc7}$ & $H_{ibc}$ & $c_{0}^{\,(I)}$ &  $3$ \\
\hline
\hline
$1$ & $G_{aibjck7}$ & $F_{aibjck}$ & $a_{0}$ &  $1$ \\
\hline
\hline
$U_{1} \, U_{2} \, U_{3} $ & non-geometric & $-F_{(0)}\,\,\,$ (Romans' mass) & $a_{3}$ & $1$ \\
\hline
\end{tabular}
\end{center}
\caption{Summary of M-theory/type IIA fluxes and couplings in $W_{\textrm{M-theory}}$/$W_{\textrm{IIA}}$. The orbifold symmetries force $\,I\neq J \neq K\,$ for all the STU couplings. These symmetries also induce a natural splitting $\,\eta^{A}=(\eta^{a} \,,\, \eta^{i} \,,\, \eta^{7})\,$ where $\,a=1,3,5\,$ and $\,i=2,4,6\,$.}
\label{Table:Fluxes_M/IIA}
\end{table}

\begin{figure}[h!]
\begin{center}
\scalebox{0.6}[0.6]{
\includegraphics[keepaspectratio=true]{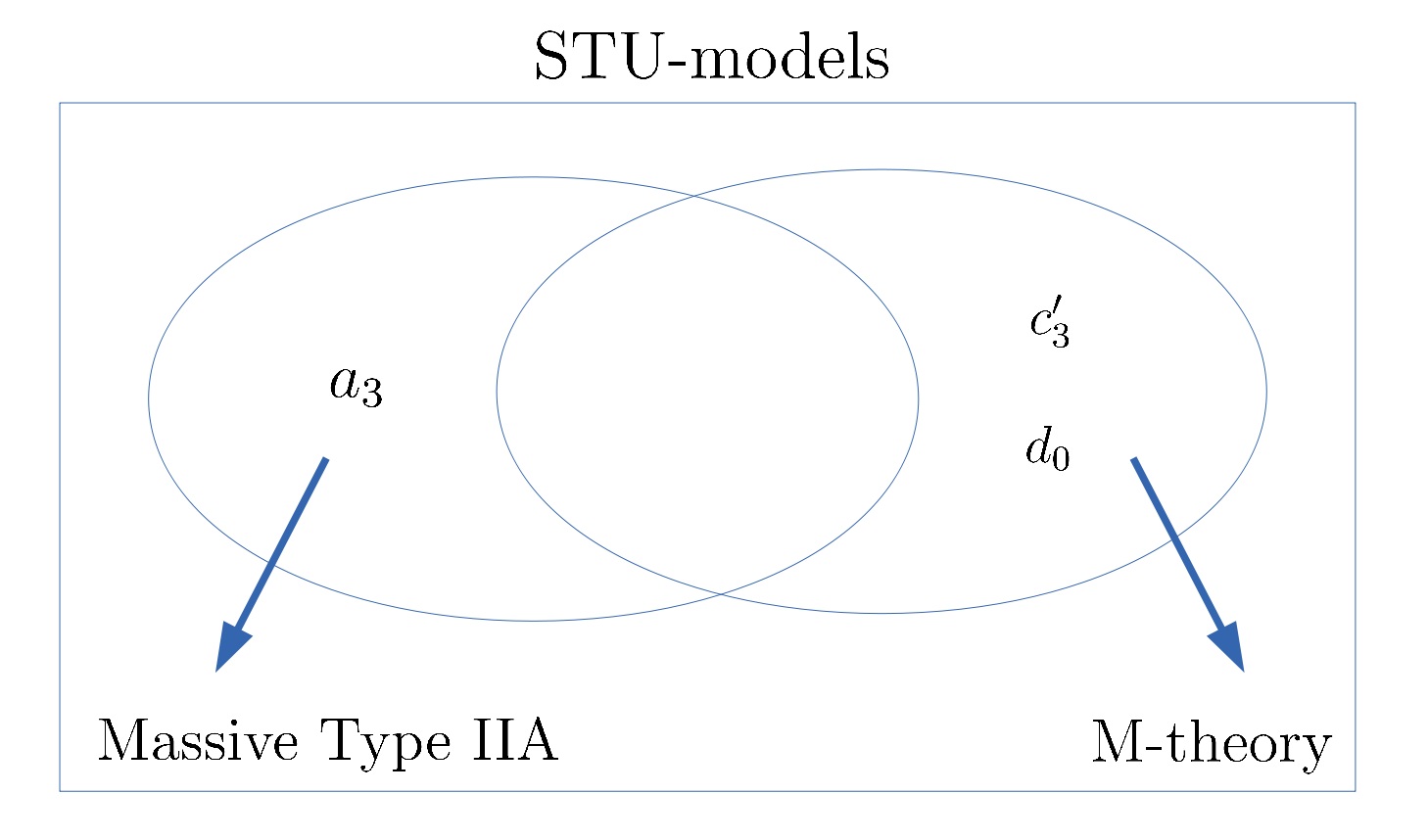}} 
\end{center}
\caption{\it{Among all possible superpotentials induced by generalised fluxes, \eqref{Superpotential_Flux_MTheory} (right side) and \eqref{Superpotential_Flux_IIA} (left side) are those ones which admit either an M-theory or massive type IIA interpretation. Whenever $a_{3}=c_{3}^{\prime}=d_{0}=0$, the corresponding STU-model falls into the large region of overlap where both the type IIA and the M-theory descriptions are available.}}
\label{fig:IIAvsM}
\end{figure}

The reductions on such geometries with fluxes have been carried out in ref~\cite{Derendinger:2004jn} (for type IIA on $X_{6}$) and refs~\cite{Dall'Agata:2005fm,Derendinger:2014wwa} (for M-theory on $X_{7}$). This paper follows the conventions of ref.~\cite{Derendinger:2014wwa} and we refer the reader to the original literature in order to follow the details of the reduction procedure. Upon reduction, the scalar sector of the $\mathcal{N}=1$ effective action contains seven complex fields\footnote{The orbifold ${X_{7}=T^{7}/\mathbb{Z}_{2}^3}$ has non-vanishing (untwisted) Betti numbers $b_{3}(X_{7})=7$. In the case of ${X_{6}=T^{6}/ \mathbb{Z}_{2}^2}$, one has (untwisted) Betti numbers $b^{-}_{2}(X_{6})=3$ and $b^{+}_{3}(X_{6})=4$ with the appropriate parity behaviour under the orientifold action $\sigma_{\textrm{O6$_{\parallel}$}}\,\, : \,\,\eta^{i} \rightarrow -\eta^{i}$.}, a.k.a \textit{moduli}, which serve as coordinates in the coset space $\left(\textrm{SL}(2)/\textrm{SO}(2)\right)^{7}$.  We denote them $\,T_{A}\,=\,(S\,,\,T_{I} \, ,\, U_{I})\,$ with $\,A=1,...,7\,$ and $\,I=1,2,3\,$. The set of moduli $\,T_{A}\,$ is the natural one to describe M-theory reductions on $X_{7}$ where one has the expansion
\be
\label{C3+Phi_expansion}
\begin{array}{ccll}
C_{(3)} + i \Phi_{(3)} &=&  \displaystyle\sum_{A=1}^{7} T_{A} \,\, \omega_{A}(y) 
\hspace{10mm} \textrm{ with } \hspace{10mm}
\omega_{A}(y) \in H^{3}(X_{7})
\end{array}
\ee
for the complexified three-form entering the superpotential (\ref{Potential_M-theory}). On the other hand, splitting the moduli as $S$, $T_{I}$ and $U_{I}$ makes the connection to the Type IIA forms entering the superpotential (\ref{Potential_IIA}) more transparent. These are given by
\be
\label{J_Omega_expansion}
\begin{array}{lllll}
J_{c} &=& \displaystyle\sum_{I=1}^{3} U_{I} \,\, \omega_{I}(y)  \hspace{10mm} & \textrm{ with }&  \hspace{10mm} \omega_{I}(y) \in H_{-}^{2}(X_{6}) \\
\Omega_{c} &=& S \,\, \alpha_{0}(y) \,+\,  \displaystyle\sum_{I=1}^{3} T_{I} \,\, \beta^{I}(y)  \hspace{10mm} & \textrm{ with }&  \hspace{10mm} \alpha_{0}(y) \,,\, \beta^{I}(y) \in H_{+}^{3}(X_{6}) \ .\\
\end{array}
\ee
Plugging the moduli expansions in (\ref{C3+Phi_expansion}) and (\ref{J_Omega_expansion}) into the M-theory and Type IIA superpotential in (\ref{Potential_M-theory}) and (\ref{Potential_IIA}), and using the background fluxes displayed in Table~\ref{Table:Fluxes_M/IIA}, one finds the M-theory result
\be
\label{Superpotential_Flux_MTheory}
\begin{array}{llll}
\mathcal{W}_{\textrm{M-theory}} &=& a_{0}  - b_{0} \, S +\displaystyle\sum_{K=1}^{3} c_{0}^{\,(K)} T_{K}  -\displaystyle\sum_{K=1}^{3} a_{1}^{\,(K)}\,U_{K} & \\[2mm]
&+& \displaystyle\sum_{K=1}^{3} a_{2}^{\,(K)} \dfrac{U_{1}U_{2}U_{3}}{U_{K}} + \displaystyle\sum_{I,J=1}^{3}  U_{I} \, \mathcal{C}_{1}^{\,(I J)}\,T_{J} +S \displaystyle\sum_{K=1}^{3} b_{1}^{\,(K)}\,U_{K} \\[2mm]
&-& \displaystyle\sum_{K=1}^{3} c_{3}'^{\,(K)} \dfrac{T_{1}T_{2}T_{3}}{T_{K}} - S \displaystyle\sum_{K=1}^{3} d_{0}^{\,(K)}\,T_{K} \ ,
\end{array} 
\ee
as well as the Type IIA result
\be
\label{Superpotential_Flux_IIA}
\begin{array}{llll}
\mathcal{W}_{\textrm{IIA}} &=& a_{0}  - b_{0} \, S + \displaystyle\sum_{K=1}^{3} c_{0}^{\,(K)} T_{K}  -\displaystyle\sum_{K=1}^{3} a_{1}^{\,(K)}\,U_{K} & \\[2mm]
&+& \displaystyle\sum_{K=1}^{3} a_{2}^{\,(K)} \dfrac{U_{1}U_{2}U_{3}}{U_{K}} + \displaystyle\sum_{I,J=1}^{3}  U_{I} \, \mathcal{C}_{1}^{\,(I J)}\,T_{J} +S \displaystyle\sum_{K=1}^{3} b_{1}^{\,(K)}\,U_{K} \\[2mm]
&-& a_{3} \, U_{1} U_{2} U_{3}  \ ,
\end{array} 
\ee
previously derived in refs\cite{Dall'Agata:2005fm,Derendinger:2014wwa}. Notice that both superpotentials only differ in the flux-induced terms appearing in the last lines: the $\,(c_{3}'^{(I)},d^{(I)}_{0})\,$ fluxes in M-theory versus the Romans' mass $\,a_{3}\,$ in type IIA. The former correspond to non-geometric fluxes in a type IIA picture and viceversa. This situation is depicted and further clarified in Figure~\ref{fig:IIAvsM}. 
Those in equations \eqref{Superpotential_Flux_MTheory} and \eqref{Superpotential_Flux_IIA} respectively, are the M-theory/Type IIA flux-induced superpotentials we will investigate in the paper.

Finally the kinetic Lagrangian for the moduli fields -- we use conventions where axions are associated to $\textrm{Re}(T_{A})$ and dilatons to $\textrm{Im}(T_{A})$ -- follows from the K\"ahler potential
\be
\label{Kaehler_STU}
\begin{array}{lll}
K &=& \displaystyle \sum_{I=A}^{7}{\log\left(-i\,(T_{A}-\bar{T}_{A})\right)} \\
&=&  -\log\left(-i\,(S-\bar{S})\right)\,-\,\displaystyle\sum_{I=1}^{3}{\log\left(-i\,(T_{I}-\bar{T}_{I})\right)}\,-\,\displaystyle\sum_{I=1}^{3}{\log\left(-i\,(U_{I}-\bar{U}_{I})\right)}\ .
\end{array}
\ee
Using (\ref{Superpotential_Flux_MTheory}) and (\ref{Superpotential_Flux_IIA}) as well as (\ref{Kaehler_STU}), the M-theory/Type IIA scalar potential can be computed from the standard $\mathcal{N}=1$ supergravity formula
\be
\label{V_N=1}
V\,=\,e^{K}\left(-3\,|\mathcal{W}|^{2}\,+\,K^{A\bar{B}}\,D_{A} \mathcal{W} \, D_{\bar{B}}\bar{\mathcal{W}}\right)\ ,
\ee
where $K^{A\bar{B}}$ is the inverse K\"ahler metric and $D_{A}\mathcal{W}=\partial_{A}\mathcal{W} + (\partial_{A}K) \mathcal{W}$ denotes the K\"ahler derivative.

\subsection{Co-calibrated G$_{2}$-structure from M-theory on $X_{7}=T^{7}/\mathbb{Z}_{2}^{3}$}

In this section we work out the co-calibrated G$_{2}$-structure associated to the orbifold space $\,{X_{7}=T^{7}/\mathbb{Z}_{2}^3}\,$ in terms of the flux-induced torsion classes. We will show that the co-calibrated G$_{2}$-structure holds regardless of the Jacobi constraints for the metric fluxes, namely, irrespective of the introduction of KK6-monopoles. This motivates the use of the G-structure as a powerful tool to uplift backgrounds with sources. In the last section, we will present the lifting to $11d$ of backgrounds with an arbitrary configuration of KK6-monopoles.

\subsubsection{M-theory metric fluxes and torsion classes}

Let us start by introducing the G$_{2}$ invariant forms for the orbifold $X_{7}=T^{7}/\mathbb{Z}_{2}^{3}$. For this particular geometry, they can be written as
\be
\label{Phi(v,J,Omega)}
\Phi_{(3)} = e^{-\frac{2}{3} \phi} \, J \wedge v + e^{-\phi} \, \Omega_{R}
\hspace{5mm} \textrm{ and } \hspace{5mm}
\star_{7d}\Phi_{(3)} = e^{-\frac{4}{3} \phi} \, J \wedge J + e^{-\phi} \, \Omega_{I} \wedge v \,
\ee
in terms of the seven-dimensional real forms $v$ (one-form) and $J$ (two-form) and the complex $\Omega=\Omega_{R} + i \, \Omega_{I}$ (three-form). These are given by 
\beq
\label{v,J,Omega-forms}
\begin{array}{llll}
v & = &  e^{\frac{2}{3} \phi} \, \eta^{7}     & , \\[2mm]
J & = &  k_{1} \,\, \eta^{1} \wedge \eta^{2} + k_{2} \,\, \eta^{3} \wedge \eta^{4} + k_{3} \,\, \eta^{5} \wedge \eta^{6}     & , \\[2mm]
\Omega & = &  \kappa \,\,  (\eta^{1} + i \, \tau_{1} \, \eta^{2}) \,\,\wedge \,\,  (\eta^{3} + i \, \tau_{2} \, \eta^{4}) \,\,\wedge \,\,  (\eta^{5} + i \, \tau_{3} \, \eta^{6}) & , 
\end{array}
\eeq
with $\kappa = \sqrt{ \frac{k_{1}k_{2}k_{3}}{\tau_{1}\tau_{2}\tau_{3}}}$, and are manifestly invariant under $\textrm{SU}(3) \subset \textrm{G}_{2} \subset\textrm{SO}(7)$. As a result, the forms $v$, $J$ and $\Omega$ specify an SU(3)-structure in $X_{7}$. However this SU(3)-structure is \textit{restricted} in the sense that it is \textit{liftable} to a G$_{2}$-structure\footnote{An SU(3)-structure in a $7d$ manifold will in general \textit{not} be liftable to a G$_{2}$-structure.}. 

Using the standard expression (\ref{g7_metric}) to obtain the $7d$ metric $g_{AB}$ in terms of $\Phi_{(3)}$ in (\ref{Phi(v,J,Omega)}), one finds
\be
\label{metric7d}
ds_{(7)}^{2} = e^{-\frac{2}{3} \phi} \, ds_{(6)}^{2} + v^{2} ,
\ee
so the $7d$ metric takes the form of a circle fibration over a $6d$ metric
\be
\label{metric6d}
ds_{(6)}^{2} = \displaystyle \sum_{I=1}^{3} \left( \frac{k_{I}}{\tau_{I}} \, (\eta^{I})^{2} \,\, +\,\,   k_{I} \, \tau_{I} \, (\eta^{I+1})^{2}  \right) \ .
\ee
The fibration in (\ref{metric7d}) ensures that the metric on the $6d$ base is in the string frame when moving to a type IIA description of the backgrounds \cite{Kaste:2002xs,Kaste:2003dh}.

The orbifold $X_{7}=T^{7}/\mathbb{Z}_{2}^{3}$ has (untwisted) Betti numbers $b_{1}(X_{7})=b_{5}(X_{7})=0$ which translates into a vanishing of the torsion classes $\tau_{1}=\tau_{2}=0$. The G$_{2}$-structure specified by the relations (\ref{dPhi}) is then called co-calibrated and takes the simple form
\be
\label{dPhiX7}
\begin{array}{lclc}
d\Phi_{3} & = & \tau_{0} \,\, \star_{7d} \Phi_{(3)} \,\, + \,\,\star_{7d} \tau_{3}  & , \\[2mm]
d\star_{7d} \Phi_{(3)} & = & 0 & .
\end{array}
\ee
The above relations can be inverted to obtain the torsion classes $\tau_{0}$ and $\tau_{3}$:
\beq
\label{Torsion_M-theory_03}
\tau_{0} =\frac{1}{7} \,  d\Phi_{(3)} \lrcorner \star_{7} \Phi_{(3)}  
\hspace{5mm} \textrm{ and } \hspace{5mm} 
\tau_{3} = \star_{7}d\Phi_{(3)} - \tau_{0} \,  \Phi_{(3)} \ . 
\eeq
An explicit computation of the torsion classes $\tau_{0}$ and $\tau_{3}$ as a function of the M-theory metric fluxes $\,{\omega_{BC}}^{A}\,$ entering the Maurer-Cartan relation  (\ref{Maurer-Cartan}) gives the following results. The torsion class $\tau_{0}$ reads
\beq
\label{tau0}
\begin{array}{llll}
\tau_{0} & = & \frac{2}{7} \, e^{\frac{1}{3}\phi} \, \left[ \kappa^{-1} \, \displaystyle \sum_{J,K=1}^{3} \mathcal{C}_{1}^{(JK)} \, \dfrac{k_{J}}{\tau_{K}} 
\,+\,  \dfrac{\kappa}{k_{1}k_{2}k_{3}} \, \displaystyle \sum_{K=1}^{3} b_{1}^{(K)} \, k_{K} \right] & \\[6mm]
&+&  \frac{2}{7} \, e^{\frac{4}{3}\phi} \, \displaystyle \sum_{K=1}^{3} a_{2}^{(K)} \, \dfrac{1}{k_{K}} \,- \,  \tfrac{2}{7} \, e^{-\frac{2}{3}\phi} \, \displaystyle \sum_{K=1}^{3} \left( {d_{0}}^{(K)} \, \dfrac{1}{\tau_{K}} + {c_{3}'}^{(K)} \, \tau_{K} \right) \ ,
\end{array}
\eeq
and is sourced by all the M-theory metric fluxes in Table~\ref{Table:Fluxes_M/IIA} with different $e^{\phi}$-weights. The second torsion class $\tau_{3}$ is a three-form which has an expansion
\beq
\label{tau3_expansion}
\tau_{3} = \tau_{3}^{(0)} \,\, \alpha_{0}  \,\,+\,\, \tau_{3 \, (I)} \,\, \beta^{I} \,\,+\,\, \tau_{3}^{(I)} \,\, \omega_{I} \ ,
\eeq
in terms of the seven basis elements of $H^{3}(X_{7})$ in (\ref{3-form X7}). The component associated to the $\alpha_{0}$ basis element in (\ref{tau3_expansion}) reads
\beq
\label{tau30}
\begin{array}{llll}
\tau_{3}^{(0)}  & = & e^{-\frac{2}{3} \, \phi} \, \left[ -\frac{2}{7} \, \displaystyle \sum_{J,K=1}^{3} \mathcal{C}_{1}^{(JK)} \, \dfrac{k_{J}}{\tau_{K}} 
\,+\, \tfrac{5}{7} \,  \dfrac{\kappa^{2}}{k_{1}k_{2}k_{3}} \, \displaystyle \sum_{K=1}^{3} b_{1}^{(K)} \, k_{K} \right] & \\[6mm]
&-&  \frac{2}{7} \, \kappa \, e^{\frac{1}{3} \, \phi} \, \displaystyle \sum_{K=1}^{3} a_{2}^{(K)} \, \dfrac{1}{k_{K}} \,-\,  \kappa \, e^{-\frac{5}{3} \, \phi} \, \displaystyle \sum_{K=1}^{3} \left( \tfrac{5}{7} \, {d_{0}}^{(K)} \, \dfrac{1}{\tau_{K}} - \tfrac{2}{7} \, {c_{3}'}^{(K)} \, \tau_{K} \right) \ ,
\end{array}
\eeq
providing again different $e^{\phi}$-weights to different fluxes. The three components associated wirth the $\beta^I$ basis elements can be written in a compact form as
\beq
\label{tau3_I}
\begin{array}{llll}
\tau_{3\,(I)}  & = & e^{-\frac{2}{3} \, \phi} \, \dfrac{1}{\tau_{I}} \, \left[  \, \tau_{1}  \tau_{2}  \tau_{3} \, \displaystyle \sum_{J,K=1}^{3} (\tfrac{2}{7}-\delta_{I}^{K}) \, \mathcal{C}_{1}^{(JK)} \, \dfrac{k_{J}}{\tau_{K}} \,+\, \tfrac{2}{7} \,  \displaystyle \sum_{K=1}^{3} b_{1}^{(K)} \, k_{K} \right] & \\[6mm]
&+&  \frac{2}{7} \, \kappa^{-1} \, e^{\frac{1}{3} \phi} \,\, \dfrac{k_{1}k_{2}k_{3}}{\tau_{I}} \, \displaystyle \sum_{K=1}^{3} a_{2}^{(K)} \, \dfrac{1}{k_{K}} & \\[6mm]
&+&  \kappa^{-1} \, e^{-\frac{5}{3} \phi} \,\, \dfrac{k_{1}k_{2}k_{3}}{\tau_{I}} \,\displaystyle \sum_{K=1}^{3} \left( (\delta_{I}^{K}-\tfrac{2}{7}) \, {d_{0}}^{(K)} \, \dfrac{1}{\tau_{K}} + (\tfrac{5}{7}-\delta_{I}^{K}) \, {c_{3}'}^{(K)} \, \tau_{K} \right) \ .
\end{array}
\eeq
Finally, the three components associated to the $\omega_{I}$ basis elements in (\ref{tau3_expansion}) can also be collectively given as
\beq
\label{tau3^I}
\begin{array}{llll}
\tau_{3}^{(I)}  & = & \kappa^{-1} \, e^{\frac{1}{3} \phi} \, k_{I} \, \left[   \displaystyle \sum_{J,K=1}^{3} (\delta_{I}^{J}-\tfrac{2}{7}) \, \mathcal{C}_{1}^{(JK)} \, \dfrac{k_{J}}{\tau_{K}} \,+\, \frac{1}{\tau_{1}  \tau_{2}  \tau_{3}}  \displaystyle \sum_{K=1}^{3} (\delta_{I}^{K}-\tfrac{2}{7})  \, b_{1}^{(K)} \, k_{K} \right] & \\[6mm]
&+&  e^{\frac{4}{3} \phi} \,\, k_{I} \, \displaystyle \sum_{K=1}^{3} (\tfrac{5}{7}-\delta_{I}^{K})  \, a_{2}^{(K)} \, \dfrac{1}{k_{K}} & \\[6mm]
&+& \frac{2}{7} \, e^{-\frac{2}{3} \phi} \,\, k_{I} \,\displaystyle \sum_{K=1}^{3} \left(  {d_{0}}^{(K)} \, \dfrac{1}{\tau_{K}} + {c_{3}'}^{(K)} \, \tau_{K} \right) \ .
\end{array}
\eeq
Notice that also the triplets $\tau_{3\,(I)}$ and $\tau_{3}^{(I)}$ come out with different $e^{\phi}$-weights for different fluxes. In summary, the above set of torsion components completely codifies the G$_{2}$-structure induced by an M-theory metric flux $\,{\omega_{BC}}^{A}\neq 0$.

\subsubsection{The matching of the scalar potentials}

Equipped with the torsion classes computed in the previous section we can move to compute the Ricci scalar using (\ref{R7}), which, in the case of a co-calibrated G$_{2}$-structure, simplifies to 
\beq
\label{R7_cocalibrated}
\mathcal{R}^{(7)}  =   \frac{21}{8} \, \tau_{0}^{2} \, - \, \frac{1}{2} \, |\tau_{3}|^{2} \ .
\eeq
We are not displaying the expression for $\mathcal{R}^{(7)}$ after plugging in the results for $\tau_{0}$ and $\tau_{3}$ since we do not gain any additional understanding on the M-theory reduction. However, let us discuss in more detail the connection to the scalar potential derived from the the $\mathcal{N}=1$ superpotential in (\ref{Superpotential_Flux_MTheory}). More concretely, we are interested in the relation (\ref{VRicci_M-theory}) reading
\beq
\label{VG2-structure-case}
V_{\textrm{G$_{2}$}} = -\frac{1}{\sqrt{g^{(7)}}} \, \mathcal{R}^{(7)} = - \frac{e^{\frac{4}{3} \phi}}{k_{1} k_{2} k_{3}} \, \left(  \frac{21}{8} \, \tau_{0}^{2} \, - \, \frac{1}{2} \, |\tau_{3}|^{2} \right) \ ,
\eeq
where we have used the expression for $7d$ metric $g^{(7)}$ in (\ref{metric7d}).

Considering only the terms coming from the twist $\,{\omega_{BC}}^{A}$ in the M-theory superpotential (\ref{Superpotential_Flux_MTheory}) -- these are the quadratic coupling in the second and third lines -- it is straightforward to compute their contribution to the full M-theory scalar potential. We will denote the purely metric-flux-induced contribution to the potential 
\beq
\label{VSTU-Mtheory-case}
V_{\textrm{M-theory}}^{\omega} = V_{\textrm{M-theory}} \Big|_{G_{(4)}=G_{(7)}=0} \ .
\eeq
In order to check whether the two potentials (\ref{VG2-structure-case}) and (\ref{VSTU-Mtheory-case}) do match, a precise identification between the $\mathcal{N}=1$ chiral moduli fields in (\ref{Superpotential_Flux_MTheory}) and the geometric moduli entering the G$_{2}$ invariant forms in (\ref{v,J,Omega-forms}) is required\footnote{We will restrict in this work to the case of vanishing axions, \textit{i.e.}, $\textrm{Re}(S)=\textrm{Re}(T)=\textrm{Re}(U)=0$. Switching off the axions does not imply a loss of generality since one can always make use of the corresponding real shift symmetries to transform them away at the price of keeping the set of gauge fluxes still completely general.}. This identification is given by \cite{Derendinger:2004jn,Villadoro:2005cu,Dall'Agata:2005fm}
\beq
\label{moduli_mapping}
\textrm{Im}(S) = e^{-\phi} \, \kappa 
\hspace{5mm} , \hspace{5mm} 
\textrm{Im}(T_{I}) = e^{-\phi} \, \kappa \, \tau_{J} \, \tau_{K} \,\,\,\,\,\, (I\neq J \neq K)
\hspace{5mm} , \hspace{5mm} 
\textrm{Im}(U_{I}) = k_{I}  \ ,
\eeq
with $\,I,J,K=1,2,3\,$ and where $\,\kappa = \sqrt{\frac{k_{1}k_{2}k_{3}}{\tau_{1}\tau_{2}\tau_{3}}}\,$ was already introduced in (\ref{v,J,Omega-forms}). After an exhaustive term-by-term check of the two potentials (\ref{VG2-structure-case}) and (\ref{VSTU-Mtheory-case}) one finds a perfect matching of the form
\beq
\label{Matching-G2}
V_{\textrm{M-theory}}^{\omega} = \frac{1}{16} \,\, V_{\textrm{G$_{2}$}} \ ,
\eeq
where the factor $1/16$ comes from the overall normalisation of the superpotential in (\ref{Potential_M-theory}).

At first sight, the perfect matching (\ref{Matching-G2}) between the potentials (\ref{VG2-structure-case}) and (\ref{VSTU-Mtheory-case}) might appear as something to be expected from the consistency of the M-theory reduction. However, in order to have a standard twisted torus interpretation of the reduction, one has to impose the Jacobi constraints
\be
\label{Jacobi_7d}
{\omega_{[AB}}^{F} \, {\omega_{C]F}}^{D} \, = \, 0 \ ,
\ee
which are satisfied in a group manifold reduction \cite{Scherk:1979zr}. Remarkably, the matching (\ref{Matching-G2}) works perfectly without imposing the conditions (\ref{Jacobi_7d}) at any moment in the computation. This fact suggests that the framework of G-structures could be a suitable one to uplift M-theory reductions -- via $11d$ universal Ans\"atze along the lines of refs~\cite{Danielsson:2009ff,Danielsson:2010bc,Danielsson:2011au} -- beyond twisted tori for which (\ref{Jacobi_7d}) does not hold. These types of reductions have been recently shown to produce full moduli stabilisation in AdS$_{4}$ vacua, and have also been connected to non-geometric type IIA backgrounds (upon reduction along $\eta^{7}$) including exotic branes lifting to KK6 monopoles in M-theory \cite{Derendinger:2014wwa}.

\subsection{Half-flat $\textrm{SU}(3)$-structures from Type IIA on $X_{6}=T^{6}/\mathbb{Z}_{2}^{2}$ }
\label{IIA_review}

Reductions of type IIA strings on a twisted $T^{6}/\mathbb{Z}_{2}^2$ orbifold with fluxes \textit{and} one single O$6$-plane (\textit{orientifold}) have been extensively studied in the literature. Such orientifold planes split the space-time coordinates into transverse and parallel directions as follows 
\be
\begin{array}{lcccc}
\textrm{O6$_{\parallel}$-plane} \, : &  &  & \underbrace{\times \, \vert \, \times \, \times \, \times}_{D = 4} \,  \, \underbrace{\times \, - \, \times \, - \, \times \, -}_{d = 6} & ,
\end{array}
\notag
\ee
and can be located at the fixed points of the $\mathbb{Z}_{2}$ involution
\beq
\label{Orientifold_action}
\sigma_{\textrm{O6$_{\parallel}$}}\, : \,\eta^{i} \rightarrow -\eta^{i} \ .
\eeq
The six-dimensional coordinates $y^{m}$ on $X_{6}$ split into orientifold-even $y^{a}$ (${a=1,3,5}$) and orientifold-odd $y^{i}$ ($i=2,4,6$) sets under (\ref{Orientifold_action}), as introduced in Table~\ref{Table:Fluxes_M/IIA}. 

We will show that the $\,\mathcal{N}=1\,$ effective STU-models arising from type IIA orientifolds of $\,X_{6}=T^{6}/\mathbb{Z}_{2}^{2}\,$ nicely fit within the framework of half-flat SU(3)-structure manifolds regardless of the Jacobi constraints for the metric fluxes. As for the previous M-theory case, what we will eventually find is a linear relation between type IIA metric flux components dressed up with the moduli and torsion classes. This will be shown explicitly in the case of \emph{vanishing axions}. In this case $\,\Omega_{R}\,$ and $\,\Omega_{I}\,$ acquire a definite parity under the orientifold involution $\sigma_{\textrm{O6$_{\parallel}$}}$ such that $\,\Omega \, \overset{\sigma}{\rightarrow} \, \bar{\Omega}$.

\subsubsection{Type IIA metric fluxes and torsion classes}
\label{sec:dictionary}

The symmetries of the $\,X_{6}=T^{6}/\mathbb{Z}_{2}^2\,$ orbifold naturally induce an SU(3)-structure on $X_{6}$ specified by an invariant two-form $J$ and a three-form $\Omega$ given by
\beq
\label{J,Omega-forms}
\begin{array}{llll}
J & = &  k_{1} \,\, \eta^{1} \wedge \eta^{2} + k_{2} \,\, \eta^{3} \wedge \eta^{4} + k_{3} \,\, \eta^{5} \wedge \eta^{6}     & , \\[2mm]
\Omega & = &  \kappa \,\,  (\eta^{1} + i \, \tau_{1} \, \eta^{2}) \,\,\wedge \,\,  (\eta^{3} + i \, \tau_{2} \, \eta^{4}) \,\,\wedge \,\,  (\eta^{5} + i \, \tau_{3} \, \eta^{6}) & ,
\end{array}
\eeq
where $\,\kappa = \sqrt{\frac{k_{1}k_{2}k_{3}}{\tau_{1}\tau_{2}\tau_{3}}}\,$ as previously introduced in (\ref{v,J,Omega-forms}). Notice that $J$ and $\Omega$ in (\ref{J,Omega-forms}) correspond to two- and three-forms in six dimensions, unlike in (\ref{v,J,Omega-forms}) where they were understood as forms in seven dimensions. It is immediate to check that they satisfy the orthogonality and normalisation conditions
\be
\begin{array}{lccclc}
\Omega \, \wedge \, J \, = \, 0 &  & \textrm{and} & & \Omega \, \wedge \, \bar{\Omega} \,=\, -\frac{4}{3} i \, J \, \wedge \, J \, \wedge \, J & . 
\end{array}
\ee

The $X_{6}$ orbifold symmetries are not compatible with the existence of one-forms (nor five-forms) thus setting $W_4=W_5=0$ in (\ref{dJdOmega}). Moreover, as a consequence of the definite $\sigma_{\textrm{O6$_{\parallel}$}}$-parity of the real and imaginary parts of $\,\Omega=\Omega_{R} + i \, \Omega_{I}\,$ in (\ref{J,Omega-forms}), the equations \eqref{dJdOmega} now take the simpler form
\be
\label{dJdOmegaHalfFlat}
\begin{array}{lclc}
dJ & = & \dfrac{3}{2} \, W_{1}\,\Omega_{I} \ + \ W_{3} & , \\[2mm]
d\Omega_{R} & = & W_{1} \, J \, \wedge \, J \ + \ W_{2} \, \wedge \, J & , \\[2mm]
d\Omega_{I} & = & 0 & ,
\end{array}
\ee
thus determining $dJ$ and $d\Omega$ purely in terms of a \emph{real} $W_{1}$ and $W_{2}$. Such an $\textrm{SU}(3)$-structure is usually referred to as \emph{half-flat} structure \cite{Ali:2006gd}. The above set of relations (\ref{dJdOmegaHalfFlat}) can again be inverted to obtain the torsion classes as a function of $J$ and $\Omega$. This process gives
\beq
\label{SU3-Torsion_classes}
W_{1} = - \frac{1}{6} \, \star_{6d} (dJ \wedge \Omega_{R})
\hspace{5mm} , \hspace{5mm}
W_{2} = -  \star_{6d} d\Omega_{R} + 2 \, W_{1} J
\hspace{5mm} , \hspace{5mm}
W_{3} = dJ - \frac{3}{2} \, W_1 \,  \Omega_{I} \ .
\eeq

Using the basis of left-invariant two- and three-forms $H^{2}(X_{6})$ and $H^{3}(X_{6})$ given in (\ref{2-form X6}) and (\ref{3-form X6}), one finds the following expansions for the torsion classes
\be
\label{W_expansions}
\begin{array}{lcclcclc}
W_{1} \ = \ w_{1} \ , & & W_{2} \ = \ {w_{2}}^{(K)} \, \omega_{K} \ , & \textrm{and} & & W_{3} \ = \ {w_{3}}_{(0)} \, \beta^{0} \, + \, {w_{3}}^{(K)} \, \alpha_{K} & ,
\end{array}
\ee
where now, due to half-flatness, all the components in (\ref{W_expansions}) are real. Although again quite tedious, the explicit computation of the torsion classes (\ref{SU3-Torsion_classes}) is performed without surprises. It results in the following expressions for the metric-flux-induced torsion classes. The torsion class $W_{1}$ reads
\beq
\label{W1_singlet}
\begin{array}{llll}
w_{1}  &=& \frac{1}{6}  \, \left[ \kappa^{-1} \, \displaystyle \sum_{J,K=1}^{3} \mathcal{C}_{1}^{(JK)} \, \dfrac{k_{J}}{\tau_{K}} 
\,+\,  \dfrac{\kappa}{k_{1}k_{2}k_{3}} \, \displaystyle \sum_{K=1}^{3} b_{1}^{(K)} \, k_{K} \right] \ ,
\end{array}
\eeq
in agreement with the structure found in the first line of (\ref{tau0}). The three components with $\omega_{K}$ in the expansion (\ref{W_expansions}) of the $W_{2}$ torsion class are collectively given by
\beq
\label{W2_torsion}
\begin{array}{llll}
w_{2}^{(I)}  &=& \kappa^{-1} \, k_{I} \, \left[   \displaystyle \sum_{J,K=1}^{3} (\tfrac{1}{3}-\delta_{I}^{J}) \, \mathcal{C}_{1}^{(JK)} \, \dfrac{k_{J}}{\tau_{K}} \,+\, \frac{1}{\tau_{1}  \tau_{2}  \tau_{3}}  \displaystyle \sum_{K=1}^{3} (\tfrac{1}{3}-\delta_{I}^{K})  \, b_{1}^{(K)} \, k_{K} \right]  \ ,
\end{array}
\eeq
also in agreement with the structure in the first line of (\ref{tau3^I}). Finally the coefficients associated with the singlet $\beta^{0}$ and the triplet $\alpha_{K}$ of basis elements in the expansion (\ref{W_expansions}) of $W_{3}$ take the form 
\beq
\label{W3_torsion}
\begin{array}{lllll}
{w_{3}}_{(0)}  &=&   \frac{1}{4} \, \tau_{1}\tau_{2}\tau_{3} \, \displaystyle \sum_{J,K=1}^{3} \mathcal{C}_{1}^{(JK)} \, \dfrac{k_{J}}{\tau_{K}} 
\,-\, \tfrac{3}{4} \,  \displaystyle \sum_{K=1}^{3} b_{1}^{(K)} \, k_{K} & ,  \\[4mm]
w_{3}^{(I)}  &=& \tau_{I} \, \left[  \, \displaystyle \sum_{J,K=1}^{3} (-\tfrac{1}{4}+\delta_{I}^{K}) \, \mathcal{C}_{1}^{(JK)} \, \dfrac{k_{J}}{\tau_{K}} \,-\, \tfrac{1}{4} \,  \frac{1}{\tau_{1}  \tau_{2}  \tau_{3} }\, \displaystyle \sum_{K=1}^{3} b_{1}^{(K)} \, k_{K} \right] & .
\end{array}
\eeq
Notice that the basis elements $(\beta^{0},\alpha_{K})$ are complementary to the basis elements $(\alpha_{0},\beta^{K})$ in (\ref{tau3_expansion}) associated to the coefficients (\ref{tau30}) and (\ref{tau3_I}). Therefore we cannot directly compare their structures. Furthermore it is straightforward to check that the above set of torsion classes given in terms of metric fluxes and moduli fields automatically satisfy the primitivity conditions in (\ref{primitivity1}) and (\ref{primitivity2}) required by the SU(3)-structure.

\subsubsection{The matching of the scalar potentials}

The set of torsion classes we obtained in the previous section can be used to compute to Ricci scalar (\ref{R6}). In this case, it has the simpler form
\be
\label{R6_half-flat}
\mathcal{R}^{(6)} \ = \ \frac{15}{2} \, \left|W_{1}\right|^{2} \, - \, \frac{1}{2} \, \left|W_{2}\right|^{2} \, - \, \frac{1}{2} \, \left|W_{3}\right|^{2} \ .
\ee
Using the $6d$ metric in (\ref{metric6d}), which is compatible with (\ref{metric6(J,I)}) if setting $\lambda^{-1}=24 \, k_{1}k_{2}k_{3}$ in (\ref{I-involution}), one finds the relation
\beq
\label{VSU3-structure-case}
V_{\textrm{SU(3)}} =  -\frac{e^{2 \phi}}{\sqrt{g^{(6)}}} \, \mathcal{R}^{(6)}= - \frac{e^{2 \phi}}{k_{1} k_{2} k_{3}} \, \left(  \frac{15}{2} \, \left|W_{1}\right|^{2} \, - \, \frac{1}{2} \, \left|W_{2}\right|^{2} \, - \, \frac{1}{2} \, \left|W_{3}\right|^{2} \right) \ .
\eeq

We are again interested in the relation between the purely metric-flux-induced contribution to the scalar potential coming from the superpotential (\ref{Superpotential_Flux_IIA}), namely
\beq
\label{VSTU-IIA-case}
V_{\textrm{IIA}}^{\omega} = V_{\textrm{IIA}} \Big|_{H_{(3)}=F_{(0)}=F_{(2)}=F_{(4)}=F_{(6)}=0} \ ,
\eeq
and the one in (\ref{VSU3-structure-case}) built in a more geometrical way out of torsion classes. Using the moduli correspondence in (\ref{moduli_mapping}), a term-by-term check reveals again a perfect matching 
\beq
\label{Matching-SU3}
V_{\textrm{IIA}}^{\omega} = \frac{1}{16} \,\, V_{\textrm{SU(3)}} \ ,
\eeq
between the two potentials (\ref{VSU3-structure-case}) and (\ref{VSTU-IIA-case}). As in the M-theory case, the matching occurs regardless of the Jacobi constraints 
\be
\label{Jacobi_IIA}
{\omega_{[mn}}^{r} \, {\omega_{p]r}}^{q} \, = \, 0 \ ,
\ee
required to have a standard twisted torus interpretation of the reduction \cite{Scherk:1979zr}.
Therefore, the SU(3)-structure can potentially be used to lift background also including KK5-monopoles. These sources were used to build simple de Sitter vacua in refs~\cite{Silverstein:2007ac,Haque:2008jz}.

\subsection{Beyond twisted tori}
\label{sec:Beyond_twisted}

We have argued that the framework of G-structures is able to accommodate twisted reductions of M-theory/type IIA regardless of the Jacobi constraints on the twist parameters, namely,
\beq
\label{ww_not0}
{\omega_{[AB}}^{F} \, {\omega_{C]F}}^{D} \, = \, 0 \,\,\,\,\,\,\, \textrm{(M-theory)}
\hspace{10mm} \textrm{or} \hspace{10mm}
{\omega_{[mn}}^{r} \, {\omega_{p]r}}^{q} \, = \, 0 \,\,\,\,\,\,\, \textrm{(type IIA)} \ .
\eeq
In the M-theory case of $X_{7}=T^{7}/\mathbb{Z}_{2}^{3}$, the first set of conditions in (\ref{ww_not0}) amounts to require the $4d$ effective action to preserve \textit{all} the $32$ supercharges ($\mathcal{N}=8$) of the $11d$ theory \cite{Dall'Agata:2005fm,Derendinger:2014wwa}. However, in the type IIA orientifold case of $X_{6}=T^{6}/\mathbb{Z}_{2}^{2}$, the second set of conditions in (\ref{ww_not0}) is not enough to guarantee the $16$ supercharges ($\mathcal{N}=4$) of the orientifolded theory and additional metric-gauge flux conditions -- in the form of tadpole conditions -- have to be supplemented to ensure a vanishing net charge of O6/D6 sources \cite{Dall'Agata:2009gv,Dibitetto:2011gm}. These sources generically reduce the amount of supersymmetry in the effective action down to $4$ supercharges ($\mathcal{N}=1$) and are secretly taken into account by the IIA superpotential (\ref{Superpotential_Flux_IIA}) \cite{Villadoro:2005cu}.

On the other hand, a non-vanishing r.h.s. in (\ref{ww_not0}) amounts to having KK6-monopoles (M-theory) or KK5-monopoles (type IIA) in the background \cite{Villadoro:2007yq}. Upon an $11d \rightarrow 10d$ reduction, KK6-monopoles give rise to KK5-monopoles as well as to O6-planes/D6-branes and more exotic sources associated to non-geometric type IIA fluxes \cite{Villadoro:2007yq,Gaillard:2009kz,Derendinger:2014wwa}. We will discuss the higher-dimensional description of these sources later on in the paper. Now we will introduce a framework where to compare both M-theory and type IIA reductions with generic background fluxes and sources going beyond the twisted tori picture, \textit{i.e.} not restricted by the conditions (\ref{ww_not0}).

\subsubsection*{SU(3)-structures in six and seven dimension}

Manifolds with SU(3)-structure in seven and six dimensions represent the natural framework to compare M-theory reductions on $X_{7}$ and type IIA orientifolds on $X_{6}$. Expressing the \mbox{G$_{2}$-structure} of $X_{7}\,$ ``\textit{a la SU(3)}" \cite{Chiossi:2002,Dall'Agata:2003ir,Dall'Agata:2005fm,Held:2011uz} will help us to understand what is the role played by the metric fluxes in M-theory that correspond to a R-R two-form flux $F_{(2)}$ \cite{Kaste:2002xs,Kaste:2003dh,Dall'Agata:2003ir,Micu:2006ey} and to non-geometric fluxes in the type IIA picture \cite{Shelton:2005cf,Aldazabal:2006up}. 

Let us derive the SU(3)-structure of the seven-dimensional manifold ${X_{7}=T^{7}/\mathbb{Z}_{2}^3}$. It is specified in terms of the seven-dimensional invariant forms $v$ (one-form), $J$ (two-form) and $\Omega$ (three-form) introduced in (\ref{v,J,Omega-forms}). The failure of the closure of $v$, $J$ and $\Omega$ is again identified with the presence of non-trivial torsion classes in the seven-dimensional manifold $X_{7}$. An explicit computation reveals
\be
\label{dvdJdOmega}
\begin{array}{lclc}
dv & = & R_{1} & , \\[2mm]
dJ & = & \dfrac{3}{2} \, W_{1}\,\Omega_{I} \ + \ W_{3} + R_{2} \wedge v & , \\[2mm]
d\Omega_{R} & = & W_{1} \, J \, \wedge \, J \ + \ W_{2} \, \wedge \, J \  + R_{3} \wedge v & , \\[2mm]
d\Omega_{I} & = &  R_{4} \wedge v    & ,
\end{array}
\ee
where $W_{1}$, $W_{2}$ and $W_{3}$ were respectively given in (\ref{W1_singlet}), (\ref{W2_torsion}) and (\ref{W3_torsion}). We will concentrate on the contributions $R_{1}$, $R_{2}$, $R_{3}$ and $R_{4}$ in (\ref{dvdJdOmega}) as they parameterise how much does the seven-dimensional SU(3)-structure deviate from being understandable as a six-dimensional one. The piece $R_{1}$ has an expansion in terms of $H^{2}(X_{6})$ given by
\beq
\label{R1_expansion}
R_{1} = R_{1}^{(I)} \, \omega_{I}
\hspace{10mm} \textrm{ with } \hspace{10mm}
R_{1}^{(I)}  = e^{\frac{2}{3}\phi} \, a_{2}^{(I)} \ ,
\eeq
and is induced by the M-theory fluxes corresponding to the R-R two-form flux $F_{(2)}$ in the type IIA picture. It is then easy to show that $R_{2}=0$ due the orbifold symmetries. The third piece $R_{3}$ has an expansion in terms of the basis elements of $H^{3}(X_{6})$ given by
\beq
\label{R3_expansion}
R_{3} = R_{3\,(0)} \, \beta^{0} \,\,+\,\,  R_{3}^{(K)} \, \alpha_{K} \ ,
\eeq
where the coefficients read
\beq
\label{R3_torsion}
\begin{array}{lllll}
{R_{3}}_{(0)}  &=&  e^{-\frac{2}{3}\phi} \, \kappa \, \tau_{1}\tau_{2}\tau_{3} \,   \displaystyle \sum_{K=1}^{3} d_{0}^{(K)} \, \frac{1}{\tau_{K}} & ,  \\[4mm]
R_{3}^{(I)}  &=& e^{-\frac{2}{3}\phi} \, \kappa \, \left[    \, {c'_{3}}^{(I)} \, \tau_{I}^{2} \, -\tau_{I}  \, \displaystyle\sum_{K =1}^{3} {c'_{3}}^{(K)} \, \tau_{K}  \, -d_{0}^{(I)}\right]  & .
\end{array}
\eeq
Finally, the last piece $R_{4}$ has an expansion in terms of the basis elements of $H^{3}(X_{6})$ given this time by
\beq
\label{R4_expansion}
R_{4} = R_{4}^{(0)} \, \alpha_{0} \,\,+\,\,  R_{4 \, (K)} \, \beta^{K} \ ,
\eeq
with coefficients
\beq
\label{R4_torsion}
\begin{array}{lllll}
R_{4}^{(0)}  &=&  e^{-\frac{2}{3}\phi} \, \kappa \,   \displaystyle \sum_{K=1}^{3} {c'_{3}}^{(K)} \, \tau_{K} & ,  \\[4mm]
R_{4  (I)}  &=& e^{-\frac{2}{3}\phi} \, \kappa \, \tau_{1} \, \tau_{2}\, \tau_{3}  \, \left[   {d_{0}}^{(I)} \, \dfrac{1}{\tau_{I}^2 }   \,- {c'_{3}}^{(I)}  \, - \, \displaystyle\sum_{K =1}^{3} {d_{0}}^{(K)} \, \frac{1}{\tau_{I} \, \tau_{K}}  \,  \right]  & .
\end{array}
\eeq
By inspection of the above torsion classes, one finds that the $R_{1}$ as well as the $R_{3}$ and $R_{4}$ terms in (\ref{dvdJdOmega}) are sourced by the metric fluxes in M-theory which are no longer metric fluxes in type IIA picture. In particular, $dv \neq 0$ is due to the R-R flux $F_{(2)}$ whereas the $R_{3}$ and $R_{4}$ contributions entering $d\Omega$ in (\ref{dvdJdOmega}) are totally induced by the M-theory fluxes $\,(c_{3}'^{(I)},d^{(I)}_{0})\,$ which correspond to non-geometric fluxes in the type IIA picture (see Table~\ref{Table:Fluxes_M/IIA}). As a result, these types of metric fluxes in seven dimensions induce deformations in the geometry that cannot be retrieved in a six-dimensional setup and therefore look like non-geometric ingredients from a six-dimensional viewpoint.

\section{Lifting STU-models to higher dimensions}
\label{sec:upliftings}

Following the philosophy of refs~\cite{Danielsson:2009ff,Danielsson:2011au}, one can use the fundamental forms and torsion classes of G-structure manifolds in order to rewrite the M-theory (type IIA) background fluxes in Table~\ref{Table:Fluxes_M/IIA} in a way that produces by construction a well-behaved stress-energy tensor with respect to the G-structure underlying the geometry. In this section we will show explicitly how this rewriting works and subsequently revisit some known supergravity solutions inspired by M-theory \cite{Derendinger:2014wwa} and type IIA \cite{Dibitetto:2011gm} compactifications (in a bottom-up sense) and reinterpret them as G-structure reductions. 

\subsection{Isotropic STU-models}

The M-theory/type IIA four-dimensional minimal supergravities in refs~\cite{Derendinger:2014wwa,Dibitetto:2011gm} that we will uplift to $11d$/$10d$ correspond to so-called \textit{isotropic} STU-models which are further invariant under a plane-exchange-symmetry \cite{Derendinger:2004jn} identifying the chiral moduli as
\beq
\label{ModuliIso}
T_{1} = T_{2}=T_{3} \equiv T
\hspace{10mm} \textrm{and} \hspace{10mm}   
U_{1} = U_{2}=U_{3} \equiv U \ ,
\eeq
as well as the M-theory/type IIA fluxes as
\beq
\label{FluxesIso}
\begin{array}{lllllllll}
\mathcal{C}_{1}^{(IJ)}=-\tilde{c}_{1} \,\,\,\,\,\,\,(I = J) & \,\,,\,\, & a_{1}^{(I)}=a_{1} & , &  b_{1}^{(I)}=b_{1}  & \,\,,\,\, &  a_{2}^{(I)}=a_{2}  & , \\[3mm]
\mathcal{C}_{1}^{(IJ)}=\phantom{-} c_{1} \,\,\,\,\,\,\,(I \neq J) & \,\,,\,\, &  c_{0}^{(I)}=c_{0} & , & d_{0}^{(I)}=d_{0} & \,\,,\,\, &  c_{3}'^{(I)}=c_{3}'  &  . 
\end{array}
\eeq
After the simplifications in (\ref{ModuliIso}) and (\ref{FluxesIso}), the \textit{isotropic} M-theory superpotential in (\ref{Superpotential_Flux_MTheory}) takes the simpler form \cite{Derendinger:2014wwa}
\beq
\label{Superpotential-MTheory-ISO}
\begin{array}{llll}
\mathcal{W}^{\textrm{(iso)}}_{\textrm{M-theory}} &=& a_{0}  - b_{0} \, S + 3 \, c_{0} \, T  - 3 \, a_{1} \,U + 3 \, a_{2} \, U^{2} + 3 \,(2 \, c_{1} - \tilde{c}_{1}) \, U\,T  + 3 \,b_{1} \,  S \, U \\[2mm]
&-& 3 \, c_{3}' \, T^{2} - 3 \, d_{0} \,  S \, T  \ ,
\end{array} 
\eeq
whereas the type IIA superpotential in (\ref{Superpotential_Flux_IIA}) gets simplified to \cite{Derendinger:2004jn,Dibitetto:2011gm}
\beq
\label{Superpotential-IIA-ISO}
\begin{array}{rlll}
\,\,\,\,\,\,\,\mathcal{W}^{\textrm{(iso)}}_{\textrm{IIA}} &=& a_{0}  - b_{0} \, S + 3 \, c_{0} \, T  - 3 \, a_{1} \,U + 3 \, a_{2} \, U^{2} + 3 \,(2 \, c_{1} - \tilde{c}_{1}) \, U\,T  + 3 \,b_{1} \,  S \, U \\[2mm]
&-& a_{3} \, U^{3} \ .
\end{array}
\eeq
Based on the scalar potential derived from the $\mathcal{N}=1$ superpotentials (\ref{Superpotential-MTheory-ISO}) and (\ref{Superpotential-IIA-ISO}), we will uplift to eleven and ten dimensions the entire set of maximally symmetric solutions found in ref.~\cite{Derendinger:2014wwa} and ref.~\cite{Dibitetto:2011gm}, respectively.

Finally, the plane-exchange symmetry (\ref{ModuliIso}) of the STU-models amounts to identify the geometric moduli in (\ref{moduli_mapping}) as 
\beq
k_{1}=k_{2}=k_{3}\equiv k  
\hspace{10mm} \textrm{and} \hspace{10mm}
\tau_{1}=\tau_{2}=\tau_{3}\equiv \tau \ .
\eeq
These moduli determine the invariant forms of the underlying G-structure of $X_{7}$ and $X_{6}$. Combined with the simplification on the fluxes (\ref{FluxesIso}), the isotropy restriction will notably reduce the expressions for the G-structure intrinsic torsion.

\subsection{M-theory uplift of STU-models}

Let us start by introducing the bosonic part of the action of the eleven-dimensional supergravity\footnote{The last term in the action is the ordinary Chern-Simons piece of 11d supergravity, so that the $\hat{G}_{(7)}$ curvature is just accounting for a pure external four-form curvature $\hat{G}_{(4)\mu\nu\rho\sigma}$. Rewriting the 11d fields of the democratic formulation of M-theory in a $4+7$ splitting and restricting to those components which are invariant under the orbifold symmetries, the above action (\ref{action11d}) is compatible with taking $\hat{C}_{(3) A_{1} A_{2} A_{3}}$ as the dynamical gauge potential as well as $\hat{G}_{(4) \mu A_{1} A_{2} A_{3}}$, $\hat{G}_{(4) A_{1} A_{2} A_{3} A_{4}}$ and $\hat{G}_{(7) A_{1} \cdots A_{7}}$ as curvatures \cite{Dall'Agata:2005fm}.}
\beq
\label{action11d}
2\kappa_{11}^{2} \, S = \int d^{11}x \, \sqrt{-g^{(11)}} \, \left(\mathcal{R}^{(11)} - \dfrac{1}{2} |\hat{G}_{(4)}|^{2}  -  \dfrac{1}{2} |\hat{G}_{(7)}|^{2}\right)  - \dfrac{1}{6 } \, \int   \hat{C}_{(3)} \wedge \hat{G}_{(4)}  \wedge \hat{G}_{(4)} \ ,
\eeq
where $|\hat{G}_{(4)}|^{2} \ \equiv \ \frac{1}{4!} \, \hat{G}_{(4) M_{1}\cdots M_{4}} \, {\hat{G}_{(4)}}^{\phantom{(4)}M_{1}\cdots M_{4}}$ and  $|\hat{G}_{(7)}|^{2} \ \equiv \ \frac{1}{7!} \, \hat{G}_{(7) M_{1}\cdots M_{7}} \, {\hat{G}_{(7)}}^{\phantom{(7)}M_{1}\cdots M_{7}}$ with ${M=0, ..., 10}$.   From the above action, the following eleven-dimensional equations of motion (EOM) and Bianchi identities (BI) follow
\beq
\label{11dEOM}
\begin{array}{rcrclc}
\textrm{EOM for $\hat{C}_{(3)}$}:  &\hspace{10mm} & d\left(\star_{11d}\hat{G}_{(4)}\right) \ + \ \frac{1}{2} \, \hat{G}_{(4)} \wedge \hat{G}_{(4)} & = & 0 & , \\[2mm]
\textrm{BI for $\hat{C}_{(3)}$}:  & & d\hat{G}_{(4)} & = & 0 & , \\[2mm]
\textrm{Einstein equations}:  & & \mathcal{R}^{(11)}_{MN} \ - \ \frac{1}{2} \, T_{MN} & = & 0 & ,
\end{array}
\eeq
where the energy-momentum tensor $T_{MN}$ is given by
\beq
\begin{array}{llll}
T_{MN} & = & \hat{G}_{(4) M P_{1} \dots P_{3}} \, \hat{G}_{(4)N}^{\phantom{(4)N}P_{1}\dots P_{3}} \, - \, \frac{1}{3} \, g^{(11)}_{MN} \, |\hat{G}_{(4)}|^{2} & \\[2mm]
&+& \hat{G}_{(7) M P_{1} \dots P_{6}} \, \hat{G}_{(7)N}^{\phantom{(7)N}P_{1}\dots P_{6}} \, - \, \frac{2}{3} \, g^{(11)}_{MN} \, |\hat{G}_{(7)}|^{2}  & .
\end{array}
\eeq
In terms of the curvature $\hat{G}_{(7)}$, the first equation in (\ref{11dEOM}) can be more conveniently rewritten as
\beq
\label{C3EOMNew}
d\left(\star_{7d}\hat{G}_{(4)}\right) \ + \ \star_{7d}\hat{G}_{(7)} \wedge \hat{G}_{(4)}  =  0 \ .
\eeq
We will use this last form when deriving the $11d$/$4d$ correspondence of the M-theory flux models.

\subsubsection*{Isotropic torsion classes and Ricci tensor of $X_{7}$}

The Ricci tensor of the $7d$ manifold $X_{7}$ with co-calibrated G$_{2}$-structure can be expressed in terms of the skew-symmetric three-form torsion \cite{Friedrich:2003}
\beq
\label{Torsion7d}
T=\frac{1}{6} \, \tau_{0} \, \Phi_{(3)} - \tau_{3} \ .
\eeq
The torsion class $\tau_{0}$ in an isotropic background is given by
\beq
\label{tau0-ISO}
\tau_{0} = \tfrac{6}{7}  \, e^{\frac{1}{3}\phi} \, k^{-\frac{1}{2}}\, \Big[ \, (2 \, c_{1} -\tilde{c}_{1}) \, \tau^{\frac{1}{2}}   +   b_{1} \, \tau^{-\frac{3}{2}} \, \Big] +  \tfrac{6}{7}  \, e^{\frac{4}{3}\phi} \, a_{2} \, k^{-1}  -  \tfrac{6}{7}  \, e^{-\frac{2}{3}\phi} \, \left( \, d_{0} \, \tau^{-1}   + c'_{3} \, \tau \, \right)  \ ,
\eeq
whereas $\tau_{3}$ still has the decomposition into a singlet and two triplets in (\ref{tau3_expansion}). The singlet component reads
\beq
\label{tau3_0-ISO}
\begin{array}{llll}
\tau_{3}^{(0)}  &=&    e^{-\frac{2}{3}\phi} \, k \, \Big[ - \tfrac{6}{7} \, (2 \, c_{1} -\tilde{c}_{1}) \, \tau^{-1}   +  \tfrac{15}{7} \,  b_{1} \, \tau^{-3} \, \Big] &  \\[4mm]
&-&  \tfrac{6}{7}  \, e^{\frac{1}{3}\phi} \, a_{2} \, k^{\frac{1}{2}} \, \tau^{-\frac{3}{2}} -   \, e^{-\frac{5}{3}\phi} \, k^{\frac{3}{2}}\,  \left( \,  \tfrac{15}{7} \, d_{0} \, \tau^{-\frac{5}{2}}   -  \tfrac{6}{7} \,  c'_{3} \, \tau^{-\frac{1}{2}} \, \right)  & .
\end{array}
\eeq
Upon the isotropic restriction, the three components of the first triplet $\tau_{3\,(I)}$ become equal and take the form 
\beq
\label{tau3_I-ISO}
\begin{array}{llll}
\tau_{3\,(1)} = \tau_{3\,(2)} = \tau_{3\,(3)}  &=&    e^{-\frac{2}{3}\phi} \, k \, \Big[ - \tfrac{1}{7} \, (2 \, c_{1} -\tilde{c}_{1}) \, \tau   +    \tfrac{6}{7} \, b_{1} \, \tau^{-1} \, \Big] &  \\[4mm]
&+&  \tfrac{6}{7}  \, e^{\frac{1}{3}\phi} \, a_{2} \, k^{\frac{1}{2}} \, \tau^{\frac{1}{2}} +   \, e^{-\frac{5}{3}\phi} \, k^{\frac{3}{2}}\,  \left( \,  \tfrac{1}{7} \, d_{0} \, \tau^{-\frac{1}{2}}   +  \tfrac{8}{7} \,  c'_{3} \, \tau^{\frac{3}{2}} \, \right)  & ,
\end{array}
\eeq
whereas the three components in the second triplet $\tau_{3}^{(I)}$ get also identified and read 
\beq
\label{tau3^I-ISO}
\begin{array}{llll}
\tau_{3}^{(1)} = \tau_{3}^{(2)} = \tau_{3}^{(3)}  &=&   \tfrac{1}{7} \,  e^{\frac{1}{3}\phi} \, k^{\frac{1}{2}} \,\Big[  \, (2 \, c_{1} -\tilde{c}_{1})  \, \tau^{\frac{1}{2}} +    b_{1} \, \tau^{-\frac{3}{2}}\, \Big] &  \\[4mm]
&+&  \tfrac{8}{7}  \, e^{\frac{4}{3}\phi} \, a_{2}  +  \tfrac{6}{7} \, e^{-\frac{2}{3}\phi} \, k \,  \left( \,  d_{0} \, \tau^{-1}   +  c'_{3} \, \tau \, \right)  & .
\end{array}
\eeq
The torsion classes in (\ref{tau0-ISO})-(\ref{tau3^I-ISO}) exactly reproduce the results of ref.~\cite{Derendinger:2014wwa} upon insertion of the flux backgrounds compatible with the AdS$_{4}$ vacua presented there.

The expression for the Ricci tensor reads
\beq
\mathcal{R}_{AB} = \frac{1}{2} \, \left( \, \frac{1}{3!} \, (dT)_{ACDE} \, {(\star_{7d} \Phi_{(3)})_{B} }^{CDE} +  \frac{1}{2!} \, T_{ACD} \, {T_{B}}^{CD}\right) \ .
\eeq
Taking (\ref{Torsion7d}) and plugging in the torsion classes computed above, one finds a diagonal Ricci tensor
\beq
\mathcal{R}_{AB} = \textrm{diag} \left( \, r_{x} \, , \, r_{y} \, , \, r_{x} \, , \, r_{y} \, , \, r_{x} \, , \, r_{y} \, , \,  r_{7}\right) \ ,
\eeq
where the three functions $r_{x}$, $r_{y}$ and $r_{7}$ depend both on the geometric moduli $(k,\tau,\phi)$ and on the M-theory metric fluxes. The first one is given by
\beq
\begin{array}{lll}
\label{rx-M-theory}
r_{x} &=&  \frac{1}{2} \, b_1^2 \, \tau ^{-4} + \frac{1}{2} \, (2 c_1-\tilde{c}_{1})^2   \\[2mm]
&-& \frac{1}{2} \, e^{2\phi} \, a_2^2  \, k^{-1} \tau^{-1} +  e^{\phi} \, (2 c_1-\tilde{c}_{1}) \, a_2 \, k^{-\frac{1}{2}} \, \tau^{-\frac{1}{2}}  \\[2mm]
&-& e^{-\phi}  \, (2 c_1-\tilde{c}_{1}) \, c_3' \,  k^{\frac{1}{2}} \, \tau^{\frac{1}{2}} 
+ \frac{1}{2} \, e^{-2\phi}  \, k \, \left( d_0^2  \, \tau^{-3} - {c_3'}^2 \, \tau \right) - 3 \, a_2 \, c_3' \ ,
\end{array}
\eeq
whereas the second one reads
\begin{eqnarray}
\label{ry-M-theory}
r_{y} &=& - \, b_1^{2}\, \tau^{-2} + 2 \, (2 c_1-\tilde{c}_{1}) \, b_1 \nonumber \\[2mm]
&-& \frac{1}{2} \, e^{2 \phi} \,  a_2^2 \, k^{-1} \, \tau 
+ \frac{1}{2} \, e^{\phi } \, k^{-\frac{1}{2}} \, \tau^{-\frac{1}{2}} \, a_2 \,  \big[ 3 \, b_1+ (2 c_1-c_{11}) \tau ^2 \big] \\[1mm]
&-&\frac{1}{2} \, e^{-\phi } \, k^{\frac{1}{2}} \, \tau^{\frac{1}{2}}  \, \big[ 3 \, b_1 \, c_3'+(2 c_1-\tilde{c}_{1}) \left(2 d_0-c_3' \tau ^2\right) \big]
- \frac{1}{2} \, e^{-2 \phi } \, k \, \tau^{-1}  (d_0^2 - {c_3'}^2 \tau ^4 ) - 3 \, a_2 \, d_0  \nonumber \ .
\end{eqnarray}
The last function takes the form
\beq
\label{r7-M-theory}
\begin{array}{lll}
r_{7} &=& \frac{3}{2} \,  e^{4 \phi } \, a_2^2 \, k^{-2}
-  \frac{3}{2} \, e^{3 \phi } \, k^{-\frac{3}{2}}  \, \tau^{-\frac{3}{2}}  \,  a_2 \,  \left[  b_1+ (2 c_1-\tilde{c}_{1}) \tau ^2 \right] \\[2mm]
&-& \frac{3}{2} \, e^{\phi } \, k^{-\frac{1}{2}} \, \tau^{-\frac{1}{2}}  \, \big[ 3 \, b_1 \, c_3' + (2 c_1-\tilde{c}_{1} ) (2 d_0 + c_3' \tau ^2) \big]
-\frac{3}{2} \, \tau^{-2} \,  (d_0 - c_3' \, \tau ^2 )^2 \ .
\end{array}
\eeq
Let us discuss the set of components of the Ricci tensor of $X_{7}$. It is worth noticing that by switching off the M-theory metric fluxes without counterparts in Table~\ref{Table:Fluxes_M/IIA} as type IIA metric fluxes, namely $a_{2}=0$ and $(d_{0},c_{3}')=(0,0)$, one finds that $r_{7}=0$ and that the expressions for $r_{x}$ and $r_{y}$ get reduced to the first line in (\ref{rx-M-theory}) and (\ref{ry-M-theory}). Additionally turning on the M-theory metric flux $a_{2}$ that corresponds to a R-R flux $F_{(2)}$ in type IIA activates the second line in (\ref{rx-M-theory}) and (\ref{ry-M-theory}) as well as the first one in (\ref{r7-M-theory}). Finally, the M-theory metric fluxes $(d_{0},c_{3}')$ which are non-geometric in the type IIA picture are responsible for the last line in (\ref{rx-M-theory}), (\ref{ry-M-theory}) and (\ref{r7-M-theory}). Computing the Ricci scalar $\mathcal{R}^{(7)}$ by contracting with the (inverse) metric in (\ref{metric7d}) reproduces the result in (\ref{R7_cocalibrated}).

\subsubsection*{Isotropic gauge backgrounds}

Recalling the M-theory gauge fluxes in Table~\ref{Table:Fluxes_M/IIA} and particularising to the isotropic case, one obtains \textit{constant} and purely internal flux backgrounds $\,\hat{G}_{(4)}=G_{(4) A_{1} \cdots A_{4}}\,$ and $\,{\hat{G}_{(7)}=G_{(7) A_{1} \cdots A_{7}}}\,$ of the form
\beq
\label{M-theory_Flux_Backgrounds}
\begin{array}{lclcllll}
G_{(4)} & = & -a_{1} \, \displaystyle\sum_{K=1}^{3} \tilde{\omega}^{K}  \ + \  b_{0} \, \beta^{0}  \ + \ c_{0} \, \displaystyle\sum_{K=1}^{3} \alpha_{K} & \hspace{5mm} \textrm{ and } \hspace{5mm} & G_{(7)} & = & a_{0} \, \eta^{1234567} & ,
\end{array}
\eeq
where $\tilde{\omega}^{I}$, $\beta^{0}$ and $\alpha_{I}$ are the seven basis elements of $H^{4}(X_{7})$ given in (\ref{4-form X7}).  Notice that the background fluxes are constant when using the set of left-invariant forms $\left\lbrace\eta^A\right\rbrace$ as the basis for expanding forms. However, and after some algebra, it can be shown that (\ref{M-theory_Flux_Backgrounds}) can be rewritten in terms of the SU(3)-structure data of $\,X_{7}\,$ as
\beq
\label{M-theory_Flux_Ansatz}
\begin{array}{lclc}
G_{(4)} & = & g_{1}(k,\tau,\phi)  \,\,  \,J \wedge J \ + \  e^{-\frac{2}{3}\phi} \left(  \, g_{2}(k,\tau,\phi)  \,\, \Omega_{I}  \ + \ g_{3}(k,\tau,\phi)  \,\,  \,\hat{W}_{3} \right) \wedge v & , \\[2mm]
G_{(7)} & = & g_{4}(k,\tau,\phi)  \,\,  \,\textrm{vol}_{7} & ,
\end{array}
\eeq
with $\,\textrm{vol}_{7}=\sqrt{g^{(7)}}\,$ and $\,\hat{W}_{3}\equiv W_{3}/|W_{3}|\,$ being the normalised version of the $W_{3}$ torsion class (\ref{W3_torsion}), and where
\beq
\label{M-theory_flux_mapping_geometric}
g_{1}  =  -\dfrac{a_1}{2 \, k^2}
\hspace{4mm} , \hspace{4mm}
g_{2}  =  \dfrac{3 \, c_0 \, \tau ^2 - b_0  }{4 \, (k \, \tau )^{3/2}}
\hspace{4mm} , \hspace{4mm}
g_{3}  =  \dfrac{\sqrt{3} \, e^{\phi} \, \left(b_0+c_0 \, \tau ^2\right)}{2 \, (k \, \tau )^{3/2}}
\hspace{4mm} , \hspace{4mm}
g_{4}  =  \dfrac{a_{0} \, e^{\frac{4}{3}\phi}}{k^3} \ ,
\eeq
are functions depending on four-dimensional quantities, namely, moduli fields and flux parameters in the superpotential (\ref{Superpotential-MTheory-ISO}). 

It is worth noticing here that (\ref{M-theory_Flux_Ansatz}) is just a rewritting of the original expansion (\ref{M-theory_Flux_Backgrounds}) in which we have used G-structure data as the basis for expanding forms instead of left-invariant forms. The form of the $G_{(4)}$ flux in (\ref{M-theory_Flux_Ansatz}) was found in ref.~\cite{Dall'Agata:2005fm} to be the one connecting M-theory to type IIA backgrounds upon reduction along the 11th space-time dimension. The $\,g_{1}\,$ function then maps to a R-R four-form flux $F_{(4)}=g_{1}(k,\tau,\phi) \, J \wedge J$ whereas the quantity between parenthesis in $\,e^{-\frac{2}{3}\phi} \left(  ... \right) \wedge v\,$ does it to a NS-NS three-form flux $\,H_{(3)}=g_{2}(k,\tau,\phi)  \,\, \Omega_{I}  \ + \ g_{3}(k,\tau,\phi)  \,\,  \,\hat{W}_{3}\,$. This is also in perfect agreement with the \textit{universal Ansatz} for type IIA reductions on SU(3)-structure manifolds investigated in refs~{\cite{Danielsson:2009ff,Danielsson:2010bc,Danielsson:2011au}}. We will elaborate more on type IIA reductions in the next section.

\subsubsection*{Matching between $11d$ and $4d$ EOM}

Equipped with the results for the Ricci tensor of $X_{7}$ and the gauge backgrounds for $\hat{G}_{(4)}$ and $\hat{G}_{(7)}$, it is now possible (and tedious) to check the eleven-dimensional equations of motion. Focusing on backgrounds with vanishing axions and constant geometric moduli, \textit{i.e.} maximally symmetric solutions, we find six independent equations descending from (\ref{11dEOM}) : three of them coming from the EOM of $\hat{C}_{(3)}$ and the other three coming from the Einstein equations. The BI for $\hat{C}_{(3)}$ is automatically satisfied due to the orbifold symmetries \cite{Dall'Agata:2005fm}. 

From the EOM of $\hat{C}_{(3)}$ we obtain the equations
\begin{equation}
\label{EOM-axions}
\partial_{\textrm{Re}(S)} V^{\textrm{(iso)}}_{\textrm{M-theory}} = 0 
\hspace{8mm} , \hspace{8mm} 
\partial_{\textrm{Re}(T)} V^{\textrm{(iso)}}_{\textrm{M-theory}} = 0
\hspace{8mm} , \hspace{8mm} 
\partial_{\textrm{Re}(U)} V^{\textrm{(iso)}}_{\textrm{M-theory}} = 0  \ ,
\end{equation}
whereas the Einstein equations reduce to
\begin{equation}
\label{EOM-dilatons}
\begin{array}{rrr}
\textrm{Im}(S) \,  \partial_{\textrm{Im}}(S) V^{\textrm{(iso)}}_{\textrm{M-theory}} &=& 0 \ ,\\[3mm]
\frac{1}{3} \, \textrm{Im}(S) \,  \partial_{\textrm{Im}(S)} V^{\textrm{(iso)}}_{\textrm{M-theory}} - \textrm{Im}(T) \,  \partial_{\textrm{Im}(T)} V^{\textrm{(iso)}}_{\textrm{M-theory}} &=& 0 \ , \\[3mm]
\frac{1}{3} \, \textrm{Im}(S) \,  \partial_{\textrm{Im}(S)} V^{\textrm{(iso)}}_{\textrm{M-theory}} + \textrm{Im}(T) \,  \partial_{\textrm{Im}(T)} V^{\textrm{(iso)}}_{\textrm{M-theory}} -2 \, \textrm{Im}(U) \,  \partial_{\textrm{Im}(U)} V^{\textrm{(iso)}}_{\textrm{M-theory}} &=& 0 \ .
\end{array}
\end{equation} 
The scalar potential $V^{\textrm{(iso)}}_{\textrm{M-theory}}$ is the six-field potential computed from the superpotential (\ref{Superpotential-MTheory-ISO}) using the standard $\,\mathcal{N}=1\,$ formula in (\ref{V_N=1}). However, both sets of equations (\ref{EOM-axions}) and (\ref{EOM-dilatons}) are understood to be evaluated at vanishing four-dimensional axions, namely, $\,\textrm{Re}(S)=\textrm{Re}(T)=\textrm{Re}(U)=0$.  As expected, the EOM for the axions descend from the EOM of the gauge potential $\hat{C}_{(3)}$ and those for the four-dimensional dilatons descend from the Einstein equations. Finally, the scalar potential $\,V^{\textrm{(iso)}}_{\textrm{M-theory}}\,$ at vanishing axions matches
\begin{equation}
V^{\textrm{(iso)}}_{\textrm{M-theory}} \Big|_{\textrm{Re}(S)=\textrm{Re}(T)=\textrm{Re}(U)=0}= -\frac{1}{16} \, \frac{1}{\sqrt{g_{(7)}}} \left(  \mathcal{R}^{(7)} - \frac{1}{2} \, |G_{(4)}|^{2} - \frac{1}{2} \, |G_{(7)}|^{2} \right) \ .
\end{equation}

As a summary, we have provided the eleven-dimensional uplift of any maximally symmetric solution (at vanishing axions) of the $\mathcal{N}=1$ superpotential (\ref{Superpotential-MTheory-ISO}) coming from twisted reductions of M-theory on an $\,{X_{7}=T^{7}/\mathbb{Z}_{2}^{3}}\,$ orbifold with background fluxes. We have shown that $X_{7}$ corresponds to a \mbox{G$_{2}$-structure} manifold whether or not the Jacobi constraints in (\ref{ww_not0}) hold. For those cases in which they don't, the underlying \mbox{G$_{2}$-structure} geometry accounts for KK6-monopoles in the background and supersymmetry is generically broken by the sources from $\mathcal{N}=8$ down to $\mathcal{N}=1$ \cite{Villadoro:2007yq,Derendinger:2014wwa}.

\subsection{Type IIA uplift of STU-models}

Let us introduce the bosonic part of \textit{massive} type IIA supergravity in ten dimensions. In the string frame, it is given by
\beq
\label{action_IIA}
\begin{array}{lcl}
2\kappa_{10}^{2} \, S_{\textrm{IIA}} & = & \displaystyle\int d^{10}x \, \sqrt{-g^{(10)}} \, \left[e^{-2\phi}\Big(\mathcal{R}^{(10)} \, + \, 4  (\partial\phi)^{2}
\, - \, \frac{1}{2} |\hat{H}_{(3)}|^{2} \Big) \, - \, \frac{1}{2} \sum\limits_{p=0,2,4} |\hat{F}_{(p)}|^{2}\right] \\[2mm]
&  + &  S_{\textrm{top.}} \, + \, S_{\textrm{loc.}}(j) \ ,
\end{array}
\eeq
where $\,|\hat{H}_{(3)}|^{2} \ \equiv \ \frac{1}{3!} \, \hat{H}_{(3)MNP}{\hat{H}_{(3)}}^{MNP}\,$ and $\,|\hat{F}_{(p)}|^{2} \ \equiv \ \frac{1}{p!} \, \hat{F}_{(p)M_{1}\dots M_{p}}{\hat{F}_{(p)}}^{M_{1}\dots M_{p}}$ with $M=0,...,9$. The above action contains a topological term of the form
\beq
\begin{array}{lcl}
S_{\textrm{top.}} & = & -\frac{1}{2} \, \displaystyle\int \left(\hat{B}_{(2)} \wedge d\hat{C}_{(3)} \wedge d\hat{C}_{(3)} \, - \, 
\tfrac{1}{3} \hat{F}_{(0)} \wedge \hat{B}_{(2)} \wedge \hat{B}_{(2)} \wedge \hat{B}_{(2)} \wedge d\hat{C}_{(3)} \right. \\[2mm]
& + & \left.\frac{1}{20} \hat{F}_{(0)} \wedge \hat{F}_{(0)} \wedge \hat{B}_{(2)} \wedge \hat{B}_{(2)} \wedge \hat{B}_{(2)} \wedge \hat{B}_{(2)} \wedge \hat{B}_{(2)} \right) \ ,
\end{array}
\eeq
and an extra piece accounting for the localised O6/D6 sources 
\beq
\begin{array}{lcl}
S_{\textrm{loc.}}(j) & = & - \displaystyle\int_{\textrm{O6/D6}} e^{-\phi} \, j_{(3)} \wedge \textrm{vol}_{4} \ ,
\end{array}
\eeq
with $j_{(3)}$ representing the $3$-form current associated with local D6/O6 sources. Taking these sources in the \emph{smeared} limit, the contribution from the local sources can be rewritten as
\beq
\begin{array}{lcl}
S_{\textrm{loc.}}(j) & = & - \displaystyle\int d^{10}x \, e^{-\phi} \, j_{(3)} \wedge \Omega \wedge \textrm{vol}_{4} \ ,
\end{array}
\eeq
with a $3$-form current, in the isotropic case, of the form
\beq
j_{(3)} \,\,=\,\, - \, N^{\parallel}_{6} \,  \beta^{0}  \,+\, N^{\perp}_{6} \,  \displaystyle\sum_{K=1}^{3} \alpha_{K} \,\, = \,\, \ j_{1}(k,\tau) \, \Omega_{I} \ \,+\, \ j_{2}(k,\tau) \, \hat{W}_{3} \ .
\eeq
The quantities $\,N^{\parallel}_{6}=N_{\textrm{O6$_{\parallel}$}} - N_{\textrm{D6$_{\parallel}$}}\,$ and $\,N^{\perp}_{6}=N_{\textrm{O6$_{\perp}$}} - N_{\textrm{D6$_{\perp}$}}\,$ respectively count the number of O6/D6 sources parallel and orthogonal to the orientifold directions and the functions $j_{1}(k,\tau)$ and $j_{2}(k,\tau)$ read
\begin{equation}
\label{J_current}
j_{1} \ = \ \frac{3 \, N_{6}^{\perp} \, \tau^{2} \, + \, N_{6}^{\parallel}}{4 \, (k \, \tau )^{3/2}}
\hspace{10mm} \textrm{and} \hspace{10mm} 
j_{2} \ = \ \frac{\sqrt{3}\, ( -N_{6}^{\parallel} \, + \, N_{6}^{\perp} \, \tau^{2})}{2 \, (k \, \tau )^{3/2}} \ .
\end{equation}

From the above action \eqref{action_IIA}, and in the smeared limit, one derives the following set of ten-dimensional EOM and BI for backgrounds with a \textit{constant} dilaton $\phi$. The EOM's for $\hat{B}_{(2)}$, $\hat{C}_{(1)}$ and $\hat{C}_{(3)}$ are respectively given by
\beq
\label{10d_EOM_gauge}
\begin{array}{rrclc}
e^{-2\phi} \, d\left(\star_{10d}\hat{H}_{(3)}\right) \ - \ \frac{1}{2} \hat{F}_{(4)} \wedge \hat{F}_{(4)} \ + \ \hat{F}_{(6)} \wedge \hat{F}_{(2)} \ + \ \hat{F}_{(8)} \wedge \hat{F}_{(0)}& = & 0 & , \\[2mm]
d\left(\star_{10d}\hat{F}_{(2)}\right) \ + \ \hat{F}_{(6)} \wedge \hat{H}_{(3)} & = & 0 & , \\[2mm]
d\left(\star_{10d}\hat{F}_{(4)}\right) \ - \ \hat{F}_{(4)} \wedge \hat{H}_{(3)} & = & 0 & ,
\end{array}
\eeq
whereas the one for the dilaton $\phi$ reads
\beq
\label{10d_EOM_dilaton}
\begin{array}{rrclc}
\frac{1}{2}e^{-2\phi} |\hat{H}_{(3)}|^{2} \ + \ \frac{1}{4} |\hat{F}_{(6)}|^{2} \ - \ \frac{1}{4} |\hat{F}_{(4)}|^{2} \ - \ \frac{3}{4} |\hat{F}_{(2)}|^{2} \ - \ \frac{5}{4} |\hat{F}_{(0)}|^{2} \ + \ 3 \,e^{-\phi} j_{1} & = & 0 & .
\end{array}
\eeq
Using purely internal background fluxes, which is more convenient to match results in the flux compactification literature, we will rewrite the first and third equations for the gauge potential in (\ref{10d_EOM_gauge}) as
\beq
\label{10d_EOM_gauge_II}
\begin{array}{rrclc}
e^{-2\phi} \, d\left(\star_{6d}\hat{H}_{(3)}\right) \ +\ \star_{6d}\hat{F}_{(6)} \wedge \hat{F}_{(4)} \ + \ \star_{6d}\hat{F}_{(4)} \wedge \hat{F}_{(2)} \ + \ \star_{6d}\hat{F}_{(2)} \wedge \hat{F}_{(0)}& = & 0 & , \\[2mm]
d\left(\star_{6d}\hat{F}_{(4)}\right) \ + \ \star_{6d}\hat{F}_{(6)} \wedge \hat{H}_{(3)} & = & 0 & .
\end{array}
\eeq
We will use this last form of the equations when deriving the 10d/4d correspondence of the type IIA flux models.

\noindent The ten-dimensional Einstein equations take the standard form
\beq
\label{10s_Einstein}
\begin{array}{rrclc}
\mathcal{R}^{(10)}_{MN} \ - \ \frac{1}{2} \, T_{MN} & = & 0 & ,
\end{array}
\eeq
where the symmetric energy-momentum tensor $T_{MN}$ is defined as
\begin{eqnarray}
T_{MN} & = & e^{2\phi} \,\sum\limits_{p}  \left(  \frac{p}{p!} \,  \hat{F}_{(p) M M_{1}\dots M_{p-1}}   {\hat{F}_{(p)N}}^{\phantom{(p)N}M_{1}\dots M_{p-1}} \, - \, \frac{p-1}{8} \,  g^{(10)}_{MN} \,  |\hat{F}_{(p)}|^{2}\right)  + \notag\\[2mm]
& + & \Big( \frac{1}{2} \, \hat{H}_{(3)M PQ} {\hat{H}_{(3)N}}^{\phantom{(3)M}PQ} - \frac{1}{4} \,  g^{(10)}_{MN} \,  |\hat{H}_{(3)}|^{2} \Big) \ + \ \Big(T_{MN}^{\textrm{loc.}}-\frac{1}{8} \, g^{(10)}_{MN} \,  T^{\textrm{loc.}}\Big)  \ ,
\end{eqnarray}
with the last term representing the contribution coming from sources. Focusing on the purely internal part, the contribution from the sources can be written as 
\beq
\left(T_{mn}^{\textrm{loc.}}-\frac{1}{8} \, g^{(6)}_{mn} \, T^{\textrm{loc.}}\right) \ = \ e^{\phi} 
\left(
\begin{array}{cc}
- \dfrac{ k}{\tau } \, \left(\dfrac{3 \, j_{1}}{2} + \dfrac{j_{2}}{\sqrt{3}}\right) & 0 \\
 0 & - k \, \tau \, \left(\dfrac{3\,  j_{1}}{2} - \dfrac{j_{2}}{\sqrt{3}}\right) \\
\end{array}
\right) \, \otimes \, \mathds{1}_{3} \ .
\eeq
In addition to the above EOM, the set of BI for the $\hat{B}_{(2)}$, $\hat{C}_{(1)}$ and $\hat{C}_{(3)}$ gauge potentials takes the form
\beq
\label{10d_BI}
\begin{array}{rrclc}
d\hat{H}_{(3)} & = & 0 & , \\[2mm]
d\hat{F}_{(2)} \ - \ \hat{F}_{(0)}  \hat{H}_{(3)} \ - \ j_{(3)} & = & 0 & , \\[2mm]
d\hat{F}_{(4)} \ + \ \hat{F}_{(2)}\wedge \hat{H}_{(3)} & = & 0 & .
\end{array}
\eeq
Plugging the set of type IIA fluxes in Table~\ref{Table:Fluxes_M/IIA} into the second equation in (\ref{10d_BI}) and using (\ref{J_current}) one recovers the standard flux-induced tadpoles for the local sources O6$_{\parallel}$/D6$_{\parallel}$ and O6$_{\perp}$/D6$_{\perp}$ \cite{Shelton:2005cf,Aldazabal:2006up}. In the isotropic case, the number of such sources is then given by
\be
\label{tadpoles_IIA}
\begin{array}{llllll}
N_{6}^{\parallel}  & = & 3\, b_{1} \, a_{2}  - b_{0} \, a_{3} & = &  N_{\textrm{O6$_{\parallel}$}} - N_{\textrm{D6$_{\parallel}$}} & , \\[2mm]  
N_{6}^{\perp}  & = & (2\, c_{1} - \tilde{c}_{1})  \, a_{2}  + c_{0} \, a_{3} & = &  N_{\textrm{O6$_{\perp}$}} - N_{\textrm{D6$_{\perp}$}} & .
\end{array}
\ee
The above combinations of fluxes in (\ref{tadpoles_IIA}) are the relevant ones when uplifting four-dimensional backgrounds to ten dimensions. As we will see now, the isotropic restriction also simplifies the torsion classes $W_{1}$, $W_{2}$ and $W_{3}$ specifying the half-flat SU(3)-structure in (\ref{dJdOmegaHalfFlat}).

\subsubsection*{Isotropic torsion classes and Ricci tensor of $X_{6}$}

Imposing the isotropic restriction on geometric moduli as well as on the type IIA fluxes, one finds that the torsion classes $\,W_{1}\,$ and $\,W_{2}\,$ read
\beq
\label{W12-ISO}
\begin{array}{rlll}
w_{1} &=& \tfrac{1}{2}  \,  k^{-\frac{1}{2}}\, \Big[ \, (2 \, c_{1} -\tilde{c}_{1}) \, \tau^{\frac{1}{2}}   +   b_{1} \, \tau^{-\frac{3}{2}} \, \Big] &  , \\[2mm]
w_{2}^{(1)} = w_{2}^{(2)} = w_{2}^{(3)} &=&  0 & .
\end{array}
\eeq
The torsion class $\,W_{3}\,$ decomposing into a singlet and a triplet of components in (\ref{W_expansions}) is given by
\beq
\label{W3-ISO}
\begin{array}{rlll}
w_{3\,(0)} &=&  \frac{3}{4} \,  (2 \, c_{1} -\tilde{c}_{1}) \, k \,  \tau^{2} \, - \, \frac{9}{4} \, b_{1} \, k   & , \\[2mm]
w_{3}^{(1)} = w_{3}^{(2)} = w_{3}^{(3)} &=& \frac{1}{4} \,  (2 \, c_{1} -\tilde{c}_{1}) \, k \, - \, \frac{3}{4} \, b_{1} \, k \, \tau^{-2} & .
\end{array}
\eeq
The above set of torsion classes in (\ref{W12-ISO}) and (\ref{W3-ISO}) can be used to characterise the SU(3)-structure underlying the four type IIA vacuum solutions presented in ref.~\cite{Dibitetto:2011gm} and further investigated in ref.~\cite{Dibitetto:2012ia}. The analysis of such solutions (see Table~2 of ref.~\cite{Dibitetto:2012ia}) reveals the following features: solutions with $\textsc{ID}=1$ (susy) and $\textsc{ID}=2,3$ (non-susy) are compatible with $W_{1} \neq 0$ and $W_{2} = W_{3}=0$, thus connecting to AdS$_{4}$ solutions of massive type IIA reductions on nearly-K\"ahler manifolds \cite{Behrndt:2004km,Behrndt:2004mj,Lust:2004ig,House:2005yc,Aldazabal:2007sn,Koerber:2008rx,Koerber:2010rn}. In contrast, the solution with $\textsc{ID}=4$ (non-susy) turns out to require $W_{1} \neq 0$, $W_{2} = 0$ and $W_{3} \neq 0$, thus connecting to AdS$_{4}$ solutions of massive type IIA reduced on more general half-flat manifolds with $W_{3} \neq 0$ \cite{Danielsson:2010bc,Danielsson:2011au}. This is a solution without the same geometry as the supersymmetric solution.

The Ricci tensor of the $6d$ manifold $X_{6}$ with half-flat SU(3)-structure can be recovered by setting $a_{2}=0$ and $(d_{0},c_{3}')=(0,0)$ in the expressions (\ref{rx-M-theory}) and (\ref{ry-M-theory}). This is turning off the M-theory fluxes corresponding to $F_{(2)}$ and non-geometric fluxes in the type IIA picture. In this way one finds that the expression for the Ricci tensor is
\beq
R_{mn} = \textrm{diag} \left( \, r_{x} \, , \, r_{y} \, , \, r_{x} \, , \, r_{y} \, , \, r_{x} \, , \, r_{y} \right) \ ,
\eeq
where the functions $r_{x}$ and $r_{y}$ depending on the geometric moduli $(k,\tau)$ and on the type IIA metric fluxes are given by 
\beq
\label{rx&ry-IIA}
r_{x} =  \tfrac{1}{2} \, b_1^2 \, \tau ^{-4} + \tfrac{1}{2} \, (2 c_1-\tilde{c}_{1})^2   
\hspace{10mm} \textrm{ and } \hspace{10mm}
r_{y} = - \, b_1^{2}\, \tau^{-2} + 2 \, (2 c_1-\tilde{c}_{1}) \, b_1 \ .
\eeq
Notice the identifications with the first lines in (\ref{rx-M-theory}) and (\ref{ry-M-theory}), respectively. As a check of consistency, the computation of the Ricci scalar $\mathcal{R}^{(6)}$ using the (inverse) metric in (\ref{metric6d}) matches the result in (\ref{R6_half-flat}).

\subsubsection*{Isotropic gauge backgrounds}

By considering massive type IIA gauge fluxes in Table~\ref{Table:Fluxes_M/IIA} particularised to the isotropic case, one obtains again \textit{constant} and purely internal flux backgrounds $\,\hat{H}_{(3)}=H_{mnp}\,$ and $\,{\hat{F}_{(p)}=F_{(p) m_{1} \cdots m_{p}}}\,$ of the form\footnote{Please note that R-R $F_{(6)}$ flux can be seen as the space-time filling component of $F_{(4)}$ in the traditional non-democratic formulation.}
\beq
\label{IIA_Flux_Backgrounds}
\begin{array}{c}
H_{(3)} =  b_{0} \, \beta^{0}  \ + \ c_{0} \, \displaystyle\sum_{K=1}^{3} \alpha_{K} \ ,\\[2mm]
F_{(0)} =  -a_{3} 
\hspace{3mm} , \hspace{3mm} 
F_{(2)}  =  a_{2} \, \displaystyle\sum_{K=1}^{3} \omega_{K} 
\hspace{3mm} , \hspace{3mm} 
F_{(4)} = -a_{1} \, \displaystyle\sum_{K=1}^{3} \tilde{\omega}^{K} 
\hspace{3mm} , \hspace{3mm} 
F_{(6)}  =  a_{0} \, \eta^{123456} \ ,
\end{array}
\eeq
where $\beta^{0}$ and $\alpha_{I}$ span $H^{(0,3)}(X_{6})$ and $H^{(2,1)}(X_{6})$ respectively, $\,\tilde{\omega}^{I}$ are the three basis elements of $H^{(2,2)}(X_{6})$ and $\omega_{I}$ are the three basis elements of $H^{(1,1)}(X_{6})$ given in equations (\ref{3-form X6}), (\ref{4-form X6}) and (\ref{2-form X6}). As happened in the M-theory case, the background fluxes are constant when using the set of left-invariant forms $\left\lbrace\eta^m\right\rbrace$ as the basis for expanding forms.  However, it is straightforward to check that (\ref{IIA_Flux_Backgrounds}) can be rewritten in terms of the SU(3)-structure data of $\,X_{6}\,$ as
\beq
\label{IIA_Flux_Ansatz}
\begin{array}{c}
H_{(3)} =  h_{1}(k,\tau)  \,\, \Omega_{I}  \ + \ h_{2}(k,\tau)  \,\,  \,\hat{W}_{3} \\[2mm]
F_{(0)} =  f_{1}(k,\tau) 
\hspace{2mm} , \hspace{2mm} 
F_{(2)}  =  f_{2}(k,\tau)\, J
\hspace{2mm} , \hspace{2mm} 
F_{(4)} = f_{3}(k,\tau)  \, \,J \wedge J
\hspace{2mm} , \hspace{2mm} 
F_{(6)}  =  f_{4}(k,\tau)  \,\textrm{vol}_{6} \ ,
\end{array}
\eeq
with $\,\textrm{vol}_{6}=\sqrt{g^{(6)}}\,$ and $\,\hat{W}_{3}\equiv W_{3}/|W_{3}|\,$ being again the normalised version of the $W_{3}$ torsion class (\ref{W3_torsion}). This form of gauge flux backgrounds was originally proposed in ref.~\cite{Danielsson:2010bc}, where it was denominated \emph{universal flux Ansatz} due to its nice feature of automatically producing a well-behaved stress-energy tensor appearing in the Einstein equations, \emph{i.e.} respecting the underlying $\textrm{SU}(3)$-structure. In our particular $X_{6}=T^{6}/\mathbb{Z}_{2}^2$ isotropic orbifold, the functions entering (\ref{IIA_Flux_Ansatz}) read
\beq
\label{IIA_flux_mapping_geometric}
\begin{array}{c}
h_{1}  =  \dfrac{ 3 \, c_0 \, \tau ^2 -b_0}{4 \, (k \, \tau )^{3/2}}  \hspace{5mm}  , \hspace{5mm} h_{2}  =  \dfrac{\sqrt{3}  \, \left(b_0+c_0 \, \tau ^2\right)}{2 \, (k \, \tau )^{3/2}} \ , \\[4mm]
f_{1} =  -a_3 
\hspace{5mm} , \hspace{5mm} 
f_{2}  =  \dfrac{a_2}{k} 
\hspace{5mm} , \hspace{5mm} 
f_{3} =-\dfrac{a_1}{2 \, k^{2}} 
\hspace{5mm} , \hspace{5mm} 
f_{4}  =  \dfrac{a_{0}}{k^3} \ ,
\end{array}
\eeq
and depend on the flux background parameters and the four-dimensional geometric moduli.

\subsubsection*{Matching between $10d$ and $4d$ EOM}

By combining the results for the Ricci tensor of $X_{6}$ with the flux \emph{Ansatz} \eqref{IIA_Flux_Ansatz}, we were able to mimic the calculation done in ref.~\cite{Danielsson:2011au} and check the ten-dimensional equations of motion. First of all, let us take a look at the BI. The one for $\hat{C}_{1}$ gives rise to the relations (\ref{J_current}) defining the current $j_{(3)}$ as a funcion of the number of O6/D6 sources in \eqref{tadpoles_IIA}. All the other BI are automatically satisfied because of the orbifold symmetries.

Focusing again on backgrounds with vanishing axions and constant geometric moduli, \textit{i.e.} maximally symmetric solutions, 
we find six independent equations descending from the ten-dimensional EOM: one of them coming from the EOM of $\hat{B}_{(2)}$, two of them from the EOM of $\hat{C}_{(3)}$, one from the EOM of $\phi$ and the last two 
coming from the Einstein equations. From the EOM of $\hat{B}_{(2)}$ we obtain the equation
\begin{equation}
\label{EOM-B2IIA}
\partial_{\textrm{Re}(U)} V^{\textrm{(iso)}}_{\textrm{IIA}} = 0  \ ,
\end{equation}
whereas from the EOM of $\hat{C}_{(3)}$ we obtain the equations
\begin{equation}
\label{EOM-C3IIA}
\partial_{\textrm{Re}(S)} V^{\textrm{(iso)}}_{\textrm{IIA}} = 0 
\hspace{8mm} , \hspace{8mm} 
\partial_{\textrm{Re}(T)} V^{\textrm{(iso)}}_{\textrm{IIA}} = 0  \ .
\end{equation}
The EOM of the ten-dimensional dilaton $\phi$ reads
\begin{equation}
\label{EOM-phiIIA}
\textrm{Im}(S) \,  \partial_{\textrm{Im}(S)} V^{\textrm{(iso)}}_{\textrm{IIA}} + \textrm{Im}(T) \,  \partial_{\textrm{Im}(T)} V^{\textrm{(iso)}}_{\textrm{IIA}} -2 \, \textrm{Im}(U) \,  \partial_{\textrm{Im}(U)} V^{\textrm{(iso)}}_{\textrm{IIA}} \, = \, 0 \ ,
\end{equation}
and the Einstein equations reduce to
\begin{equation}
\label{EOM-dilatonsIIA}
\begin{array}{rrr}
5 \, \textrm{Im}(S) \,  \partial_{\textrm{Im}(S)} V^{\textrm{(iso)}}_{\textrm{IIA}} - \frac{1}{3} \, \textrm{Im}(T) \,  \partial_{\textrm{Im}(T)} V^{\textrm{(iso)}}_{\textrm{IIA}} + \frac{2}{3} \,\textrm{Im}(U) \,  \partial_{\textrm{Im}(U)} V^{\textrm{(iso)}}_{\textrm{IIA}}&=& 0 \ , \\[3mm]
3 \, \textrm{Im}(S) \,  \partial_{\textrm{Im}(S)} V^{\textrm{(iso)}}_{\textrm{IIA}} - \frac{7}{3} \,\textrm{Im}(T) \,  \partial_{\textrm{Im}(T)} V^{\textrm{(iso)}}_{\textrm{IIA}} - \frac{2}{3} \, \textrm{Im}(U) \,  \partial_{\textrm{Im}(U)} V^{\textrm{(iso)}}_{\textrm{IIA}} &=& 0 \ .
\end{array}
\end{equation} 
This time the scalar potential $V^{\textrm{(iso)}}_{\textrm{IIA}}$ is the six-field potential computed from the superpotential (\ref{Superpotential-IIA-ISO}). As in the M-theory case, the equations (\ref{EOM-B2IIA})-(\ref{EOM-dilatonsIIA}) are evaluated at vanishing axions $\,\textrm{Re}(S)=\textrm{Re}(T)=\textrm{Re}(U)=0$.  Notice that the EOM for the axions descend from the EOM of the gauge potentials $\hat{B}_{(2)}$ and $\hat{C}_{(3)}$ whereas those for the four-dimensional dilatons descend from the ten-dimensional dilaton $\phi$ and the Einstein equations. Finally, the scalar potential $\,V^{\textrm{(iso)}}_{\textrm{IIA}}\,$ at vanishing axions matches
\begin{equation}
V^{\textrm{(iso)}}_{\textrm{IIA}} \Big|_{\textrm{Re}(S)=\textrm{Re}(T)=\textrm{Re}(U)=0}= -\frac{1}{16} \, \frac{e^{2\phi}}{\sqrt{g^{(6)}}} \left(  \mathcal{R}^{(6)} - \frac{1}{2} \, |H_{(3)}|^{2} 
- \frac{1}{2} \, e^{2\phi} \, \sum\limits_{p}|F_{(p)}|^{2} + 4 \,  e^{\phi} \,  j_{1} \, \right) \ .
\end{equation}

Similarly to the M-theory case, we have now provided the ten-dimensional uplift of any maximally symmetric solution (at vanishing axions) of the $\mathcal{N}=1$ superpotential (\ref{Superpotential-IIA-ISO}) coming from twisted reductions of massive type IIA on an $\,{X_{6}=T^{6}/\mathbb{Z}_{2}^{2}}\,$ orbifold with background fluxes. We have shown that $X_{6}$ corresponds to a \mbox{SU(3)-structure} manifold whether or not the Jacobi constraints in (\ref{ww_not0}) hold. For those cases in which they don't, the underlying \mbox{SU(3)-structure} geometry accounts for KK5-monopoles in the background and supersymmetry is generically broken by the sources from $\mathcal{N}=8$ down to $\mathcal{N}=1$ \cite{Villadoro:2007yq,Derendinger:2014wwa}.

\section{Conclusions}
\label{sec:discussion}

We have investigated massive type IIA/M-theory reductions in the presence of background fluxes using the framework of G-structures and their intrinsic torsion. Taking a \textit{twisted} orbifold as the internal space for the reduction  -- the $\,X_{6}=T^{6}/\mathbb{Z}_{2}^2\,$ orbifold for type IIA and the $\,X_{7}=T^{7}/\mathbb{Z}_{2}^3\,$ orbifold for M-theory  -- we have established a precise correspondence between four dimensional data, namely, twist parameters $\,\omega\,$ (metric fluxes) and moduli fields, and the torsion classes of the flux-induced G-structure underlying the reduction.

These types of twisted orbifold reductions produce classes of minimal supergravities dubbed STU-models which are specified in terms of $\,\mathcal{N}=1\,$ flux-induced superpotentials. Remarkably, we observed that the Ricci scalar of the internal G-structure manifold computed from the flux-induced torsion classes exactly reproduces the scalar potential obtained from the flux-induced superpotentials without having to impose any extra condition on the twist parameters $\,\omega\,$. In other words, we found a perfect matching regardless of whether or not the usual Jabobi constraints $\,\omega \, \omega = 0\,$ on the twist parameters are satisfied. These Jacobi constraints are required in an ordinary Scherk-Schwarz reduction, thus taking our G-structure reductions beyond the standard twisted tori picture. 

Relaxing the Jacobi constraints has been connected to the presence of (smeared) KK-monopoles in the background \cite{Villadoro:2007yq}. Therefore, it becomes natural to propose the framework of G-structures and their intrinsic torsion as the natural playground where to describe KK-monopoles in a geometric manner. With this motivation, we have studied two different classes of STU-models and have provided their uplift to type IIA string theory with an internal $\textrm{SU}(3)$-structure manifold and to M-theory with an internal $\textrm{G}_{2}$-structure manifold. In particular, we showed that any maximally symmetric solution to the four-dimensional equations is automatically a solution also to the equations of motion in ten or eleven dimensions. The uplift nicely works without requiring the Jacobi constraints at any time in the computations. As a consequence, we can accommodate  (smeared) objects such as KK5-monopoles in type IIA string theory and KK6-monopoles in M-theory and provide a higher-dimensional description of the AdS$_{4}$ vacua presented in refs~\cite{Dibitetto:2011gm,Derendinger:2014wwa}. However, even though we can map STU-models to explicit $\textrm{SU}(3)$- or $\textrm{G}_{2}$-structures, it still does not mean that we in general know the actual geometry that realises the torsion classes. This is a problem that needs to be examined on a case by case basis.
 
Our uplifting formulas assume the sources to be smeared. From the point of view of type IIA string theory in ten dimensions this can be criticised. In general, at least when supersymmetry is broken, one can expect considerable differences between localised and smeared solutions. With localisation there are warp factors to be taken into account that will modify the effective four-dimensional dynamics. It is not even clear whether the solutions of the smeared models will have anything to do with solutions of the localised ones. The same criticism cannot be voiced against the M-theory scenario. There, everything is geometry and all of the $4d$ physics can be realised in the form of smooth configurations of geometry and gauge fields. The way KK-monopoles are captured by our construction is a nice illustration of how this may happen. In this context it is interesting to speculate that the smeared solutions in type IIA string theory actually do have a physical meaning – even though one must go beyond IIA supergravity and use an M-theory perspective to make sense of them. The relation between string theory and M-theory would be like the one between classical physics and quantum mechanics.  It would be interesting to investigate this perspective further.

%
%

\section*{Acknowledgments}

We would like to thank G. Dall'Agata, J.-P. Derendinger, G. Inverso, L. Martucci and D. Tsimpis for stimulating and interesting discussions. And especially J.-P. Derendinger for discussions that led to an initial collaboration on this work. The work of UD and GD was supported by the Swedish Research Council (VR). The work of AG was supported by the Swiss National Science Fundation and by the ERC Advanced Grant no. 246974, ``Supersymmetry: a window to non-perturbative physics''. AG wants to thank the COST MP1210 action for support and also the Department of Physics and Astronomy in Uppsala for its hospitality during the early stages of this work.

\newpage

%
%

\appendix

\section{Geometrical data of $X_{7}=T^{7}/\mathbb{Z}_{2}^{3}\,$ and $X_{6}=T^{6}/\mathbb{Z}_{2}^2$}
\label{sec:app1}

In this appendix we summarise the geometry of the orbifolds $\,X_{7}=T^{7}/\mathbb{Z}_{2}^{3}\,$ and $\,X_{6}=T^{6}/\mathbb{Z}_{2}^2\,$ which are extensively investigated in this paper.

\subsubsection*{The orbifold $X_{7}=T^{7}/\mathbb{Z}_{2}^{3}\,$}

To describe the geometry of $X_{7}=T^{7}/\mathbb{Z}_{2}^{3}\,$ we will adopt the conventions of ref.~\cite{Derendinger:2014wwa}. The orbifold symmetries are defined via the actions
\be
\label{T7-orbifold-action}
\begin{array}{lclclc}
\theta_{1} & : & \left(y^{1},\,y^{2},\,y^{3},\,y^{4},\,y^{5},\,y^{6},\,y^{7}\right) & \longmapsto & \left(-y^{1},\,-y^{2},\,-y^{3},\,-y^{4},\,y^{5},\,y^{6},\,y^{7}\right) & , \\[1mm]
\theta_{2} & : & \left(y^{1},\,y^{2},\,y^{3},\,y^{4},\,y^{5},\,y^{6},\,y^{7}\right) & \longmapsto & \left(-y^{1},\,-y^{2},\,y^{3},\,y^{4},\,-y^{5},\,-y^{6},\,y^{7}\right) & , \\[1mm]
\theta_{3} & : & \left(y^{1},\,y^{2},\,y^{3},\,y^{4},\,y^{5},\,y^{6},\,y^{7}\right) & \longmapsto & \left(y^{1},\,-y^{2},\,y^{3},\,-y^{4},\,y^{5},\,-y^{6},\,-y^{7}\right) & . 
\end{array}
\ee
The above orbifold symmetries induce a natural splitting of the $\eta^{A=1,...7}$ coordinates 
\beq
\eta^{A} \rightarrow (\,\eta^{a} \,,\, \eta^{i} \,,\, \eta^{7}\,)
\eeq
with $a=1,3,5$ and $i=2,4,6$. It is then easy to verify that there are only three- and four-forms which are invariant under the action of (\ref{T7-orbifold-action}). This translates into the ${X_{7}=T^{7}/\mathbb{Z}_{2}^{3}}$ orbifold having non-vanishing (untwisted) Betti numbers $b_{3}(X_{7})=b_{4}(X_{7})=7$. 

The corresponding basis for $H^{3}(X_{7})$ and $H^{4}(X_{7})$ can be explicitly given. The $H^{3}(X_{7})$ basis is spanned by the three-forms\footnote{We introduce here the short-hand $\eta^{A_{1} \dots A_{p}} \, \equiv \, \eta^{A_{1}} \, \wedge \, \cdots \, \wedge \, \eta^{A_{p}}$.}
\beq
\label{3-form X7}
\begin{array}{llllllll}
\omega_{1}=\eta^{12} \wedge \eta^{7} &\, , \,& \omega_{2}=\eta^{34} \wedge \eta^{7} &\, , \,& \omega_{3}=\eta^{56} \wedge \eta^{7} &\, , \,&  &  \\[2mm]
\alpha_{0}=\eta^{135} &\, , \,& \beta^{1}=\eta^{146}  &\, , \,& \beta^{2}=\eta^{362} &\, , \,& \beta^{3}=\eta^{524} &  ,
\end{array}
\eeq
whereas that of $H^{4}(X_{7})$ consists of the four-form basis elements
\beq
\label{4-form X7}
\begin{array}{llllllll}
\tilde{\omega}^{1}=\eta^{3456} &\, , \,& \tilde{\omega}^{2}=\eta^{1256} &\, , \,& \tilde{\omega}^{3}=\eta^{1234}  &\, , \,& \\[2mm] 
\beta^{0}=\eta^{246} \wedge \eta^{7}  &\, , \,& \alpha_{1}=\eta^{235} \wedge \eta^{7}  &\, , \,& \alpha_{2}=\eta^{451} \wedge \eta^{7} &\, , \,& \alpha_{3}=\eta^{613} \wedge \eta^{7} & .
\end{array}
\eeq
The two basis satisfy the orthogonality conditions
\beq
\label{orghogonality-X7}
\int_{X_{7}} \omega_{I} \wedge \tilde{\omega}^{J} =  \mathcal{V}_{7} \, \delta_{I}^{J}
\hspace{5mm} , \hspace{5mm}
\int_{X_{7}} \alpha_{0} \wedge \beta^{0} =  - \mathcal{V}_{7} 
\hspace{5mm} , \hspace{5mm}
\int_{X_{7}} \beta^{I} \wedge \alpha_{J} =  -\mathcal{V}_{7} \, \delta^{I}_{J} \ ,
\eeq
with $\,I, J=1,2,3\,$. The volume and orientation of $X_{7}$ are then defined as $\mathcal{V}_{7}=\int_{X_{7}} \eta^{1234567}$.

\subsubsection*{The orbifold $X_{6}=T^{6}/\mathbb{Z}_{2}^{2}\,$}

Let us now consider the $\mathbb{Z}_{2}\,\times\,\mathbb{Z}_{2}$ group given by $\left\{1, \, \theta_{1}, \, \theta_{2}, \, \theta_{1}\theta_{2}\right\}$ acting only on the coordinates  $\eta^{m=1,...,6} \, \equiv \, \left(\eta^{a}, \, \eta^{i}\right)$ spanning the $T^{6}$ in $T^{7}=T^{6} \times S^{1}$. Using the set of 1-forms $\eta^{m}$ on the internal manifold $\,X_{6}\,\equiv\,T^{6}/\mathbb{Z}_{2}^2\,$ one can construct the following set of invariant 2-forms
\be
\label{2-form X6}
\omega_{1}=\eta^{12} 
\hspace{5mm} , \hspace{5mm}
\omega_{2}=\eta^{34}
\hspace{5mm} , \hspace{5mm}
\omega_{3}=\eta^{56} \ ,
\ee
spanning $H^{(1,1)}(X_{6})$ with $\,h^{(1,1)}=3$. Correspondingly, this defines the following set of 4-forms
\be
\label{4-form X6}
\tilde{\omega}^{1}=\eta^{3456} 
\hspace{5mm} , \hspace{5mm}
\tilde{\omega}^{2}=\eta^{1256}
\hspace{5mm} , \hspace{5mm}
\tilde{\omega}^{3}=\eta^{1234} \ ,
\ee
spanning $H^{(2,2)}(X_{6})$ with $\,h^{(2,2)}=3$. The set of invariant 3-forms is, instead, decomposed as
\be
\Lambda^{3}(X_{6}) \ = \ H^{(3,0)}(X_{6}) \ \oplus \ H^{(2,1)}(X_{6}) \ \oplus \ \ H^{(1,2)}(X_{6}) \ \oplus \ H^{(0,3)}(X_{6}) \ ,
\ee 
which are spanned by
\begin{equation}
\label{3-form X6}
\begin{array}{llllllll}
\alpha_{0} = \eta^{135} & , & \alpha_{1}=\eta^{235} & , & \alpha_{2}=\eta^{451} & , & \alpha_{3}=\eta^{613} & , \\[2mm]
\beta^{0} = \eta^{246} & , & \beta^{1}=\eta^{146} & , & \beta^{2}=\eta^{362} & , & \beta^{3}=\eta^{524} & ,
\end{array}
\end{equation}
where $h^{(2,1)} \, = \, h^{(1,2)} \, = \, 3$. The above invariant forms on $X_{6}$ satisfy the following orthonormality relations
\be
\begin{array}{lclclc}
\int\limits_{X_{6}}\omega_{K} \, \wedge \, \tilde{\omega}^{J} \, = \, \mathcal{V}_{6} \,\delta^{J}_{K} & , &
\int\limits_{X_{6}}\alpha_{0} \, \wedge \, \beta^{0} \, = \, -\mathcal{V}_{6} \,\delta^{J}_{K} & , & 
\int\limits_{X_{6}}\alpha_{K} \, \wedge \, \beta^{J} \, = \, \mathcal{V}_{6} \,\delta^{J}_{K} & ,
\end{array}
\ee
where the volume and orientation of $X_{6}$ are defined as $\mathcal{V}_{6} \, \equiv \, \int\limits_{X_{6}} \eta^{123456}$.

%
%

\small


\bibliography{references}
\bibliographystyle{utphys}

\end{document}